\documentclass[epjc3]{svjour3}

\RequirePackage[T1]{fontenc}

\smartqed 

\RequirePackage{graphicx}
\RequirePackage{mathptmx}      
\RequirePackage{flushend}
\RequirePackage[numbers,sort&compress]{natbib}
\usepackage{amsfonts,amsmath,amssymb}

\usepackage{caption}
\usepackage{subcaption}
\usepackage[english]{babel}
\usepackage{array,multirow} 
\usepackage{color}
\usepackage{bm} 
\usepackage{inputenc}
\DeclareMathOperator{\Tr}{Tr}
\providecommand{\ZZ}{\mathbb{Z}}

\journalname{Eur. Phys. J. C}

\begin{document}

\title{Vacuum stability conditions of the economical $3-3-1$ model from copositivity}

\thankstext{e1}{e-mail: bruce.sanchez@ufabc.edu.br}
\thankstext{e2}{e-mail: rgambini@ifi.unicamp.br}
\thankstext{e3}{e-mail: calvarez@ifi.unicamp.br}
\institute{Centro de Ci\^encias Naturais e Humanas, Universidade Federal do ABC - UFABC,\\09.210-170, Santo Andr\'e, SP, Brazil\label{addr1} 
\and Instituto de F\'isica Gleb Wataghin - UNICAMP,\\13083-859, Campinas,
SP, Brazil.\label{addr2}}
\author{B. L. S\'anchez-Vega\thanksref{e1,addr1}\and Guillermo Gambini\thanksref{e2,addr2}\and C. E. Alvarez-Salazar\thanksref{e3,addr2} }

\date{Received: date / Accepted: date}

\maketitle

\begin{abstract}
By applying copositivity criterion to the scalar potential of the economical $3-3-1$ model, we derive necessary and sufficient bounded-from-below conditions at tree level. Although these are a large number of intricate inequalities for the dimensionless parameters of the scalar potential, we present general enlightening relations in this work. Additionally, we use constraints coming from the minimization of the scalar potential by means of the orbit space method, the positivity of the squared masses of the extra scalars, the Higgs boson mass, the $Z'$ gauge boson mass and its mixing angle with the SM $Z$ boson in order to further restrict the parameter space of this model.
\end{abstract}

\section{Introduction\label{introduction}}
Models addressing open questions concerning the standard model of particle physics (SM) usually resort to the use of new symmetries and/or the addition of extra particles. As a first example, we can mention models implementing different see-saw mechanisms (type I, II and III) which introduce bosonic or fermionic degrees of freedom in order to explain tiny neutrino masses and their mixings \cite{PhysRevD.22.2227,DeRomeri2017,chulia2018seesaw,Bertuzzo:2018ftf}. In combination with that,  new abelian and non-abelian symmetries are also invoked in order to obtain highly predictive scenarios where not only neutrino masses are fixed but also further correlations between neutrino oscillation parameters appear \cite{Nishi:2016wki,Nishi:2018vlz,Babu2003,Gupta2012,Ferreira2012,PhysRevD.86.073007,Ding2013,PhysRevD.93.093009,Chen:2015siy,deSalas:2017kay,Capozzi:2018ubv,Reig:2016tuk}. A series of models with a matter content larger than the one of the SM are those dealing with the impressive observation that almost thirty percent of the energy content of the Universe is due to dark matter (DM)~\cite{PlanckCollaboration2018}. Arguably, the simplest model providing a DM candidate is that which extends the SM only by a real scalar transforming in a non trivial way under a stabilizing $\ZZ_2$ discrete symmetry~\cite{Silveira:1985rk,Feng:2014vea}. However, other well-motivated models based on supersymmetry~\cite{Pagels1982Supersymmetry,sanchez2013case,montero2017supersymmetric,Guio:2018puj}, extra dimensions~\cite{HOOPER200729}, the $B-L$ symmetry~\cite{LINDNER2011324,montero2011neutrino,Sanchez-Vega:2014rka,Bruce2015Fermionic} and those with an Axion/ALP~\cite{Peccei1977CP,Weinberg1978New,Wilczek1978Strong,WITTEN1984,Carvajal2015,carvajal2017linking,montero2018axion} have been widely considered (see~\cite{Bertone:2016nfn} for a review). All of them have additional symmetries and extra particles in their physical spectrum.  

In the same vein, the so-called $3-3-1$ models are interesting extensions of the SM. The fundamental idea behind all these models is to extend the $\mathrm{SU}(2)_{L}$ gauge group to the $\textrm{SU}(3)_{L}$ one,  in other words, the total gauge group of these models is $\textrm{G}_{331}\equiv\textrm{SU}(3)_{C}\otimes \textrm{SU}(3)_{L}\otimes \textrm{U}(1)_{N}$.  Here, $C$ and $L$ stand, as in the SM, for color and left chirality, respectively. However, $N$ stands for a new charge different from the SM hypercharge, $Y$, and its values are assigned to obtain the latter after the first spontaneous symmetry breaking. More specifically, the values of $N$ together with an embedding parameter  $b$ determine the electric charges of the matter content in these models since the electric charge operator is $Q=T_{3}-bT_{8}+N\mathbf{1}_{3\times3}$ ~\cite{sanchez2018new}, where $T_{3,\,8}$ are the diagonal Gell-Mann matrices, $\mathbf{1}_{3\times3}$ is the $3\times3$ identity matrix and the parameter $b$ can take two values: $1/\sqrt{3}$ or $\sqrt{3}$.

In this paper, we consider the $3-3-1$ model with $b=1/\sqrt{3}$ and the simplest scalar sector as proposed in Refs. \cite{foot19943,ponce2002analysis} after a systematic study of all possible $3-3-1$ models without exotic electric charges. This model is known in the literature as the ``economical $3-3-1$ model\textquotedblright$\,$ and has some appealing features that turn it arguably the most interesting $3-3-1$ model. Among these properties we can mention that right-handed neutrinos, $N_{a}$, are in the same $\textrm{SU}(3)_{L}$ multiplet as the SM leptons, $\nu_{a}$ and $e_{a}$. This is possible because the fundamental representation of the $\textrm{SU}(3)_{L}$ gauge group is larger than the $\textrm{SU}(2)_{L}$ one and the parameter $b=1/\sqrt{3}$ allows the $\textrm{SU}(3)_{L}$ multiplet to have two electrically neutral components: $\nu_{a}$ and $N_{a}^{c}$. This property allows for massive neutrinos at tree level. Yet, agreement with experiments is reached only when one-loop contributions to neutrino masses are considered  \cite{dong2007neutrino}. Other no less important features of this model are the possibility of implementing the Peccei-Quinn mechanism in order to solve the strong CP problem \cite{montero2011natural} and the existence of axion dark matter \cite{montero2018axion}. Needless to say, this model also shares some appealing features with other versions \cite{clavelli1974conditions,lee19773,lee19783,singer19793,singer1980canonical,pisano19923,frampton1992chiral,montero1993neutral} such as the capability to shed some light on the family replication issue of the SM. 

Although the economical $3-3-1$ model has several appealing features, it also introduces a considerable number of
degrees freedom that turn it less predictive. For instance, its scalar potential has at least nineteen coupling constants. It is a large number when compared with two couplings in the SM scalar potential.  Needless to say, a large number of extra Yukawa couplings is allowed by the $3-3-1$ gauge group. Therefore, in this paper we search for constraints that allow to reduce this, in some sense, undesirable freedom. More specifically, we study vacuum stability at tree level, i.e. the conditions guaranteeing that the scalar potential is bounded from below in all directions in the field space as the field norms approach to infinity. It is well-known that in the SM, at tree level, it is enough to make the Higgs boson quartic coupling positive  \cite{Gabrielli:2013hma,Masina:2013aba,Antipin:2013sga}. Nevertheless, in the case of the economical $3-3-1$ model we face a more complicated problem even at tree level since we have to deal with nineteen coupling constants and the scalar fields belong to the fundamental and anti-fundamental representations of $\textrm{SU}(3)_{L}$. However, the problem becomes simpler when a $\ZZ_2$ symmetry acting on some  fields is considered. This symmetry is already used in the $3-3-1$ literature  \cite{sanchez2018new, foot1993lepton, montero2011natural, mizukoshi2011wimps, foot19943, cogollo2014excluding, dias2005naturally, dong2006fermion} with different motivations. In the present context, this symmetry not only reduces the number of coupling constants to fourteen, but also makes the quartic terms in the scalar potential to have a $\lambda_{ij}\phi_i^2\phi_j^2$ form. Therefore, demanding that the scalar potential is bounded from below as the field norms approach to infinity is equivalent to ensuring that the $\lambda_{ij}$ matrix is copositive (positive on nonnegative vectors) \cite{cottle1970classes,kaplan2000test,kim1982general,kannike2016vacuum}. In addition, to make the problem even more tractable, we use the method of the orbit space in Refs. \cite{abud1981geometry,abud1983geometry} which greatly reduces the number of variables. At the end, the problem of vacuum stability is reduced to study the copositivity of a $3\times3$ matrix. This provides seventeen inequalities, for the ten quartic couplings, that at first sight seem too complicated in order to provide useful analytical relations. However, combining these inequalities with constraints coming from the first and second derivative tests for a minimum of the scalar potential, we manage to find enlightening analytical constraints for these coupling constants.   

Finally, with the aim of restricting the rest of the scalar couplings we turn our attention to the scalar mass spectrum, since all of the squared scalar masses must be positive in general. However, these masses also depend on the vacuum expectation values (VEVs) of the scalar fields. Thus, we use the experimental limits on the mass of an extra neutral gauge boson  $Z'$  \cite{Aad:2015zhl, Sirunyan:2017exp,aaboud2018measurement,tanabashi2018aps}, and the bound on the $Z-Z'$ mixing angle in this model \citep{cogollo2008fermion} to estimate the VEVs. Doing so, we find relations for three of the remaining four scalar potential couplings. Also, the experimental limit on the Higgs mass \cite{Aad:2015zhl, Sirunyan:2017exp,aaboud2018measurement,tanabashi2018aps} is used to constrain even more some couplings.

The rest of the paper is organized as follows. In Sec. \ref{model} we introduce the generalities of the economical $3-3-1$ model with a $\ZZ_2$ symmetry that allows a scalar potential with quartic terms in a biquadratic form of the field norms. In Sec. \ref{vacuum}, taking advantage of this property, we search for constraints on the scalar potential couplings imposing the vacuum stability conditions at tree level. Specifically, we use the method of the orbit space to simplify the application of the first and second derivative tests together with the copositivity criterion. After finding clear and useful relations for the values of some scalar potential parameters, in Sec. \ref{scalarmasses}, we go further applying the positivity of the scalar masses and the experimental Higgs mass in order to constrain more scalar parameters. Finally, we present our conclusions in Sec. \ref{conclusions}.

\section{The model\label{model}}
In order to cancel the quantum gauge anomalies, the left-handed fermions of the economical $3-3-1$ model have to belong to the $\left(1,\mathbf{3},\,-1/3\right)$ representation for the three lepton families and to the $\left(\mathbf{3},\,\mathbf{3},\,1/3\right)$, $\left(\mathbf{3},\,\bar{\mathbf{3}},\,0\right)$ representations for the quark families. More specifically, 
\begin{align}
\textrm{Leptons: }
F_{aL}&=\left(\nu_{a},\,e_{a},\,N_{a}^{c}\right)_{L}^{\textrm{T}}\sim\left(1,\mathbf{3},\,-1/3\right), \label{eq:1}\\
\textrm{Quarks: }Q_{L} & =\left(u_{1}\nonumber,\,d_{1},\,u{}_{4}\right)_{L}^{\textrm{T}}\sim\left(\mathbf{3},\,\mathbf{3},\,1/3\right),\\
Q_{bL} & =\left(d_{b},\,u_{b},\,d_{b+2}\right)_{L}^{\textrm{T}}\sim\left(\mathbf{3},\,\bar{\mathbf{3}},\,0\right),\label{eq:2}
\end{align}
where $a=1,2,3$, $b=2,\,3$ and ``$\sim$'' means the transformation properties under the local symmetry group. Furthermore, in the right-handed field sector we have 
\begin{align}
\textrm{Leptons: }e_{aR} & \sim\left(1,\,1,\,-1\right),\label{eq:3}\\
\textrm{Quarks:\,\ }u_{sR} & \sim\left(3,\,1,\,2/3\right),\quad d_{tR}\sim\left(3,\,1,\,-1/3\right),\label{eq:4}
\end{align}
where $a$ takes the same values as in Eq. $\eqref{eq:1}$, $s=1,\dots,4$ and $t=1,\dots,5$.

It is also necessary to introduce, at least, three SU$\left(3\right)_{L}$ triplets, $\rho,\,\eta,\,\chi$, in order to generate the appropriate fermion and boson masses. Note that the scalar sector can not be additionally reduced due to the appearance of accidental symmetries implying some massless ``u'' and ``d'' type quarks at all orders in perturbation theory as shown in Ref. \cite{montero2015accidental}. In other words, the three SU$\left(3\right)_{L}$ scalar triplets are necessary to totally break the $\textrm{G}_{331}$ symmetry to the $\textrm{U}\left(1\right)_{Q}$ symmetry and, at the same time, to give the phenomenologically appropriate masses for the quarks. These scalar triplets are given by 
\begin{equation}
\label{eq:5}
\rho=\left(\rho_{1}^{+},\,\rho_{2}^{0},\,\rho_{3}^{+}\right)^{\textrm{T}}\sim\left(1,\,\mathbf{3},\,2/3\right), \quad\eta=\left(\eta_{1}^{0},\,\eta_{2}^{-},\,\eta_{3}^{0}\right)^{\textrm{T}}\sim\left(1,\:\mathbf{3},\,-1/3\right),
\end{equation} 
\begin{equation}
\label{eq:6}
\quad\chi=\left(\chi_{1}^{0},\,\chi_{2}^{-},\,\chi_{3}^{0}\right)^{\textrm{T}}\sim\left(1,\:\mathbf{3},\,-1/3\right).
\end{equation}

Once these fermionic and bosonic fields are introduced in the model, the most general Yukawa Lagrangian, renormalizable and invariant under the local gauge group, reads
\begin{equation}
\label{eq:7}
\mathcal{L}_{\textrm{Yuk}}=\mathcal{L}_{\textrm{Yuk}}^{\rho}+\mathcal{L}_{\textrm{Yuk}}^{\eta}+\mathcal{L}_{\textrm{Yuk}}^{\chi},
\end{equation}
with 
\begin{eqnarray}
\mathcal{L}_{\textrm{Yuk}}^{\rho} & = & \alpha_{t}\bar{Q}_{L}d_{tR}\rho+\alpha_{bs}\bar{Q}_{bL}u_{sR}\rho^{*}+\text{Y}_{aa^\prime}\epsilon_{ijk}\left(\bar{F}_{aL}\right)_{i}\left(F_{a^\prime L}\right)_{j}^{c}\left(\rho^{*}\right)_{k}+\textrm{Y}^\prime_{aa^\prime}\bar{F}_{aL}e_{a^\prime R}\rho\nonumber \\
 &  & +\textrm{H.c.,}\label{eq:8}\\
\mathcal{L}_{\textrm{Yuk}}^{\eta} & = & \beta_{s}\bar{Q}_{L}u_{sR}\eta+\beta{}_{bt}\bar{Q}_{bL}d_{tR}\eta^{*}+\textrm{H.c.},\label{eq:9}\\
\mathcal{L}_{\textrm{Yuk}}^{\chi} & = & \gamma_{s}\bar{Q}_{L}u_{sR}\chi+\gamma{}_{bt}\bar{Q}_{bL}d_{tR}\chi^{*}+\textrm{H.c.},\label{eq:10}
\end{eqnarray}
where $\epsilon_{ijk}$ is the Levi-Civita symbol and $a^\prime,i,j,k=1,2,3$ and $a$, $b$, $s$, $t$ are in the same range as in Eqs. (\ref{eq:1}-\ref{eq:4}). It is also straightforward to write down the most general scalar potential consistent with gauge invariance and renormalizability as 
\begin{eqnarray}
\label{eq:11}
V\left(\eta,\rho,\chi\right) & = & -\mu_{1}^{2}\eta^{\dagger}\eta-\mu_{2}^{2}\rho^{\dagger}\rho-\mu_{3}^{2}\chi^{\dagger}\chi-\mu_{4}^{2}\chi^{\dagger}\eta \nonumber\\
 &  & +\lambda_{1}\left(\eta^{\dagger}\eta\right)^{2}+\lambda_{2}\left(\rho^{\dagger}\rho\right)^{2}+\lambda_{3}\left(\chi^{\dagger}\chi\right)^{2}+\lambda_{4}\left(\chi^{\dagger}\chi\right)\left(\eta^{\dagger}\eta\right)\nonumber \\
 &  & +\lambda_{5}\left(\chi^{\dagger}\chi\right)\left(\rho^{\dagger}\rho\right)+\lambda_{6}\left(\eta^{\dagger}\eta\right)\left(\rho^{\dagger}\rho\right)+\lambda_{7}\left(\chi^{\dagger}\eta\right)\left(\eta^{\dagger}\chi\right)\nonumber \\
 &  & +\lambda_{8}\left(\chi^{\dagger}\rho\right)\left(\rho^{\dagger}\chi\right)+\lambda_{9}\left(\eta^{\dagger}\rho\right)\left(\rho^{\dagger}\eta\right)+\lambda_{10}\left(\chi^{\dagger}\eta\right)^{2}\nonumber\\
 &  & +\lambda_{11}\left(\chi^{\dagger}\eta\right)\left(\eta^{\dagger}\eta\right)+\lambda_{12}\left(\chi^{\dagger}\eta\right)\left(\chi^{\dagger}\chi\right)+\lambda_{13}\left(\chi^{\dagger}\eta\right)\left(\rho^{\dagger}\rho\right)\nonumber \\
 &  & +\lambda_{14}\left(\chi^{\dagger}\rho\right)\left(\rho^{\dagger}\eta\right)-\frac{\lambda_{15}}{\sqrt{2}}\epsilon_{ijk}\eta_{i}\rho_{j}\chi_{k}+\textrm{H.c.}\,.
\end{eqnarray}
From Eqs. \eqref{eq:5} and \eqref{eq:6}, we can see that there are five electrically neutral scalars, $\rho_{2}^{0},$ $\eta_{1}^{0},$ $\,\eta_{3}^{0},$ $\chi_{1}^{0},$ $\chi_{3}^{0}\,$ and, in principle, all of them can gain VEVs. However, it is well known that the minimal vacuum structure needed to give masses for all the particles in the model is 
\begin{equation}
\label{eq:12}
\left\langle \rho\right\rangle =\frac{1}{\sqrt{2}}\left(0,\,v_{\rho},\,0\right)^{\textrm{T}},\,\,\left\langle \eta\right\rangle =\frac{1}{\sqrt{2}}\left(v_{\eta},\,0,\,0\right)^{\textrm{T}},\,\,\left\langle \chi\right\rangle =\frac{1}{\sqrt{2}}\left(0,\,0,\,v_{\chi}\right)^{\textrm{T}},
\end{equation}
which correctly reduces the $\textrm{G}_{331}$ symmetry to the $\textrm{U}\left(1\right)_{Q}$ one if $v_{\eta}^2+v_{\rho}^2\equiv v^2=246^2$ GeV$^2$. There is at least another reason for choosing the minimal vacuum structure given in Eq. \eqref{eq:12}. If the remaining neutral scalars, $\eta_{3}^{0}$ and $\chi_{1}^{0}$, also gain VEVs, dangerous Nambu-Goldstone bosons can arise in the physical spectrum, as shown in Ref. \cite{sanchez2018new}. Therefore, in order to avoid this issue and looking for simplicity, in this paper we are going to consider only the minimal vacuum structure given in Eq. \eqref{eq:12}.

We also consider the model with an extra simplifying assumption that is quite common in the $3-3-1$ literature \cite{sanchez2018new, foot1993lepton, montero2011natural, mizukoshi2011wimps, foot19943, cogollo2014excluding, dias2005naturally, dong2006fermion}. It consists on the imposition of a discrete $\ZZ_2$ symmetry given by: $\chi\rightarrow-\chi$, $u_{4R}\rightarrow-u{}_{4R}$, $\,d{}_{\left(4,5\right)R}\rightarrow-d{}_{\left(4,5\right)R}$ and the other fields being even under $\ZZ_2$. This symmetry not only brings simplicity to the model, allowing, for instance, to interpret the $\chi$ scalar as the responsible for the first step in the symmetry breaking pattern, but also to mitigate the FCNC issues \cite{GomezDumm:1994tz}. Moreover, with this $\ZZ_{2}$ symmetry the PQ mechanism can be easily implemented \cite{montero2011natural} and in some cases dark matter candidates can be stabilized \cite{montero2018axion,cogollo2014excluding}. In this scenario, the Yukawa Lagrangian interactions given in Eqs.  
(\ref{eq:8}-\ref{eq:10}) are slightly modified to
\begin{eqnarray} 
\mathcal{L}_{\textrm{Yuk},\,\mathbb{Z}_{2}}^{\rho} & = & \alpha_{a}\bar{Q}_{L}d_{aR}\rho+\alpha_{ba}\bar{Q}_{bL}u_{aR}\rho^{*}+\text{Y}_{aa'}\varepsilon_{ijk}\left(\bar{F}_{aL}\right)_{i}\left(F_{bL}\right)_{j}^{c}\left(\rho^{*}\right)_{k}+\text{Y}'_{aa'}\bar{F}_{aL}e_{a'R}\rho+\nonumber \\
 &  & \textrm{H.c.,}\label{eq:13}\\
\mathcal{L}_{\textrm{Yuk},\,\mathbb{Z}_{2}}^{\eta} & = & \beta_{a}\bar{Q}_{L}u_{aR}\eta+\beta{}_{ba}\bar{Q}_{bL}d_{aR}\eta^{*}+\textrm{H.c.},\label{eq:14}\\
\mathcal{L}_{\textrm{Yuk},\,\mathbb{Z}_{2}}^{\chi} & = & \gamma_{4}\bar{Q}_{L}u_{4R}\chi+\gamma{}_{b\left(b+2\right)}\bar{Q}_{bL}d_{\left(b+2\right)R}\chi^{*}+\textrm{H.c.}\,.\label{eq:15}
\end{eqnarray}
Note that in this case the $\chi$ triplet only couples to the $u_{4R}$ and $d_{(4,5)R}$ quarks, and it is also the reason to interpret $\chi$ as the scalar responsible for the first step in the symmetry breaking pattern. Furthermore, after imposing the $\ZZ_2$ symmetry, the scalar potential becomes 
\begin{eqnarray}
\label{eq:16}
V\left(\eta,\rho,\chi\right) & = & V_{\mathbb{Z}_{2}}\left(\eta,\rho,\chi\right) + V_{\mathrm{Soft}} \left(\eta,\rho,\chi\right), \nonumber \\
 & = & -\mu_{1}^{2}\eta^{\dagger}\eta-\mu_{2}^{2}\rho^{\dagger}\rho-\mu_{3}^{2}\chi^{\dagger}\chi \nonumber\\
 &  & +\lambda_{1}\left(\eta^{\dagger}\eta\right)^{2}+\lambda_{2}\left(\rho^{\dagger}\rho\right)^{2}+\lambda_{3}\left(\chi^{\dagger}\chi\right)^{2}+\lambda_{4}\left(\chi^{\dagger}\chi\right)\left(\eta^{\dagger}\eta\right)\nonumber \\
 &  & +\lambda_{5}\left(\chi^{\dagger}\chi\right)\left(\rho^{\dagger}\rho\right)+\lambda_{6}\left(\eta^{\dagger}\eta\right)\left(\rho^{\dagger}\rho\right)+\lambda_{7}\left(\chi^{\dagger}\eta\right)\left(\eta^{\dagger}\chi\right)\nonumber \\
 &  & +\lambda_{8}\left(\chi^{\dagger}\rho\right)\left(\rho^{\dagger}\chi\right)+\lambda_{9}\left(\eta^{\dagger}\rho\right)\left(\rho^{\dagger}\eta\right)+\lambda_{10}\left(\chi^{\dagger}\eta\right)^{2}\nonumber\\
 &  & -\frac{\lambda_{15}}{\sqrt{2}}\epsilon_{ijk}\eta_{i}\rho_{j}\chi_{k}+\textrm{H.c.}\,.
\end{eqnarray}
It is important to note that, even though the term $\lambda_{15}\epsilon_{ijk}\eta_{i}\rho_{j}\chi_{k}$ softly breaks the $\ZZ_2$ symmetry, it must be included because in its absence appears a QCD axion with a small decay constant, $11.5\text{ keV}\leq f_{a}\leq 246\text{ GeV}$, already ruled out by experiments \cite{bardeen1987constraints, carvajal2017linking}. See Ref. \cite{sanchez2018new} for a detailed study of this case. 


\section{Minimization and vacuum stability\label{vacuum}}
Now, we turn our attention to find constraints on the $\mu_i-$ and $\lambda_i-$scalar parameters coming from minimization and vacuum stability. The general minimization of the scalar potential in Eq. \eqref{eq:16} is a difficult task due to the large number of free parameters in the scalar potential (14 free parameters), the large number of components of the scalars triplets in the model (18 components for the $\rho,\,\eta,\,\chi$ triplets) and the degeneracies of the extremal points of the potential required by the  $\textrm{G}_{331}-$invariance. Fortunately, there is a powerful tool to simplify this problem which consists in working with the norm of the fields and orbit parameters. This method to minimize scalar potentials, also known as the method of the orbit space, is detailed in Refs. \cite{abud1981geometry,kim1982general,abud1983geometry,kannike2016vacuum} in the context of spontaneous symmetry breaking. It has been used, for instance, in models with $\textrm{SU}(5)$ and $\textrm{SO}(10)$ gauge symmetries when scalars belong to different representations \cite{frautschi19825, kim1982n, kim1982general, kim1984orbit}. The crucial observation of the method is that working with the norm of the fields, $|\phi|$ ($|\phi|^2 \equiv \phi^{*}_{k}\phi_k$ - where a sum over repeated indices is implied) and 
the invariant orbit parameters $\boldsymbol{\theta}$'s (generically defined by ${\boldsymbol{\theta}}=\frac{f_{ijkl}\,\phi^{*}_{i}\phi_{j}\phi^{*}_k\phi_l}{|\phi|^4}$) contain all the information
needed to determine the minimum of the potential and, in addition, greatly reduce the number of variables. 

In order to apply the method of orbit space, let's define the invariant orbit parameters of the scalar potential in Eq. \eqref{eq:16} as 
\begin{eqnarray}
\label{eq:17}
\boldsymbol{\theta}_1(\hat{\eta},\hat{\chi}) =   \hat{\chi}^*_j\hat{\eta}_j\hat{\eta}^*_i\hat{\chi}_i, \quad{\boldsymbol{\theta}}_2 (\hat{\rho},\hat{\chi}) =  \hat{\chi}^*_j\hat{\rho}_j\hat{\rho}^*_i\hat{\chi}_i, \quad {\boldsymbol{\theta}}_3 (\hat{\eta},\hat{\rho}) = \hat{\eta}^*_i\hat{\rho}_i \hat{\rho}^*_j\hat{\eta}_j,\nonumber\\
\boldsymbol{\theta}_4(\hat{\eta},\hat{\chi}) =  (\hat{\chi}^*_i \hat{\eta}_i)^2+ \mathrm{H.c.}, \quad
{\boldsymbol{\theta}}_5 (\hat{\eta},\hat{\rho},\hat{\chi}) = \epsilon_{ijk}\hat{\eta}_i \hat{\rho}_j \hat{\chi}_k + \mathrm{H.c.\, ,} 
\end{eqnarray}
where $\hat{\phi}_k = \phi_k / |\phi|$ and a sum over the repeated indices is implied. Note that all directional information is contained within the $\boldsymbol{\theta}$ parameters. 

The scalar potential takes the following simple form when written using the $\boldsymbol{\theta}$ parameters:
\begin{eqnarray}
\label{eq:18}
V(\eta,\rho,\chi)&=&-\mu_1^2 |\eta |^2 -\mu_2^2 | \rho |^2 -\mu_3^2 | \chi |^2 + \lambda_1 |\eta |^4 +\lambda_2 | \rho |^4 +\lambda_3| \chi |^4  \nonumber\\ 
&+& (\lambda_4 + \lambda_7 \boldsymbol{\theta}_1 + |\lambda_{10}| \boldsymbol{\theta}_4) |\eta |^2|\chi |^2+ (\lambda_5 +\lambda_8 \boldsymbol{\theta}_2) |\rho |^2 |\chi |^2  + (\lambda_6 +\lambda_9 \boldsymbol{\theta}_3) |\eta |^2|\rho |^2  \nonumber \\
&-&\frac{|\lambda_{15}|}{\sqrt{2}} \boldsymbol{\theta}_5\, |\eta | |\rho | |\chi |,
\end{eqnarray}
where we have used the fact that making a redefinition of the scalar fields, e.g. $\eta_k  \rightarrow e^{-i\delta_{10}/4} \eta_k, \,\chi_k  \rightarrow  e^{i\delta_{10}/4}\chi_k$ and $\rho_k  \rightarrow  e^{-i\delta_{15}} \rho_k$, the phases of  $\lambda_{10}=|\lambda_{10}|$ $e^{i\delta_{10}}$ and $\lambda_{15} = |\lambda_{15}|$ {$e^{i\delta_{15}}$} couplings can be absorbed. Therefore, all couplings in the scalar potential can be considered as real numbers, without loss of generality.

Treating the modules of $\eta$, $\rho$, $\chi$, and the $\boldsymbol{\theta}$ parameters as independent variables, we could apply the first and second derivative tests to provide general conditions to have a minimum of the scalar potential in Eq. \eqref{eq:16}. However, we will restrict ourselves to the phenomenologically interesting vacuum given in Eq. \eqref{eq:12}. In other words, we consider the directional minima in the particular direction given by
\begin{eqnarray}
\label{eq:19}
(\boldsymbol{\theta}_1)_0 = 0,\hspace{0.15in} (\boldsymbol{\theta}_2)_0 = 0, \hspace{0.15in} (\boldsymbol{\theta}_3)_0 = 0,\hspace{0.15in} (\boldsymbol{\theta}_4)_0 = 0,\hspace{0.15in}(\boldsymbol{\theta}_5)_0 = 2. 
\end{eqnarray}
Thus, taking the directional derivative of the scalar potential in Eq. \eqref{eq:18} in relation to the norm of the scalar fields, we obtain an expression for the $\mu_i$ parameters:
\begin{eqnarray}
\label{eq:20}
&&  2\mu_1^2=2\lambda_1 v_{\eta}^2+\lambda_6 v_{\rho}^2+\lambda_4 v_{\chi}^2-|\lambda_{15}| v_{\rho} v_{\chi}/v_{\eta}  , \nonumber\\
&&  2\mu_2^2=2\lambda_2 v_{\eta}^2+\lambda_6 v_{\rho}^2+\lambda_5 v_{\chi}^2-|\lambda_{15}| v_{\eta} v_{\chi}/v_{\rho}, \nonumber\\
&&  2\mu_3^2=2\lambda_4 v_{\eta}^2+\lambda_5 v_{\rho}^2+\lambda_3 v_{\chi}^2-|\lambda_{15}| v_{\eta} v_{\rho}/v_{\chi}, 
\end{eqnarray}
where we have used $|\eta|_0=v_{\eta}/\sqrt{2} $,  $|\rho |_0=v_{\rho}/\sqrt{2} $ and $|\chi |_0=v_{\chi}/\sqrt{2} $ which are the norm of the fields in the vacuum direction in Eq. \eqref{eq:12}.

Additional conditions on the scalar potential parameters come from second derivative test. Specifically, all principal minors of the Hessian matrix $H_0$ evaluated at the vacuum, $(H_0)_{ij} = \frac{\partial^2 V}{\partial \phi_i \partial \phi_j} \bigg|_{\phi=\phi_0}, $  must be positive (see for example \cite{pita1991algebra, callahan2010advanced}).  A straightforward calculation gives the Hessian matrix at the directional minimum as:
\begin{equation}
\label{eq:21}
H_0 = \left(
\begin{array}{ccc}
4\lambda_1 v_{\eta}^2+|\lambda_{15}| v_{\rho} v_{\chi}/ v_{\eta} & 2 \lambda_6 v_{\eta} v_{\rho}  -|\lambda_{15}| v_{\chi}  &2\lambda_4 v_{\eta} v_{\chi}  - |\lambda_{15}| v_{\rho}  \\
 \star &  4\lambda_2 v_{\rho}^2+|\lambda_{15}| v_{\eta} v_{\chi} /v_{\rho} &  2 \lambda_5 v_{\rho} v_{\chi} -  |\lambda_{15}|v_{\eta}\\
\star & \star   &  4\lambda_3 v_{\chi}^2+|\lambda_{15}| v_{\eta} v_{\rho} /v_{\chi}
\end{array}\right),
\end{equation}
where we have used the relations in Eq. \eqref{eq:20} in order to simplify the Hessian matrix. From the positivity of the principal minors we get the conditions
\begin{eqnarray}
\label{eq:22}
 \lambda_1>-|\lambda_{15}|  v_\rho v_{\chi}/4v_{\eta}^3 ,\,\lambda_2 >-|\lambda_{15}| v_{\eta} v_{\chi}/4v_{\rho}^3,\, \lambda_3 >-|\lambda_{15}| v_{\eta} v_{\rho}/4v_{\chi}^3,&& \nonumber\\
\frac{ |\lambda_{15}| v_\rho -\sqrt{(H_0)_{11}(H_0)_{33}}}{2v_\eta v_\chi}< \lambda_4  < \frac{ |\lambda_{15}| v_\rho +\sqrt{(H_0)_{11}(H_0)_{33}}}{2v_\eta v_\chi}, &&\nonumber \\
\frac{ |\lambda_{15}|v_\eta -\sqrt{(H_0)_{22}(H_0)_{33}}}{2 v_\rho v_\chi} < \lambda_5 < \frac{ |\lambda_{15}|v_\eta +\sqrt{(H_0)_{22}(H_0)_{33}}}{2 v_\rho v_\chi}, &&\nonumber \\ 
 \frac{|\lambda_{15}| v_\chi -\sqrt{(H_0)_{11}(H_0)_{22}}}{2 v_\eta v_\rho} < \lambda_6  < \frac{|\lambda_{15}| v_\chi +\sqrt{(H_0)_{11}(H_0)_{22}}}{2 v_\eta v_\rho}, && \nonumber\\
\det\, H_0 >0, &&
\end{eqnarray}
where we have not written explicitly an analytical expression for the determinant of $H_0$, which is not very enlightening. In addition, $\det \, H_0>0$ will be automatically satisfied provided we consider the positivity of the square masses of the CP-even scalars as will be shown in the next section.

So far, we have studied the conditions for the vacuum configuration in Eq. \eqref{eq:12} to be a minimum of the scalar potential. Nevertheless, for the scalar potential in Eq. \eqref{eq:18} to make physical sense, it has to be bounded from below in the large field limit for all possible directions in the field space. To obtain these constraints it is enough to work with the quartic terms, which can be written in a matrix form as
\begin{eqnarray}
\label{eq:23}
V_4&=&\lambda_1 |\eta |^4 +\lambda_2 | \rho |^4 +\lambda_3| \chi |^4 + (\lambda_4 + \lambda_7 \boldsymbol{\theta}_1 +|\lambda_{10}| \boldsymbol{\theta}_4) |\eta |^2|\chi |^2  \nonumber \\ 
&+& (\lambda_5 +\lambda_8 \boldsymbol{\theta}_2) |\rho |^2 |\chi |^2  + (\lambda_6 +\lambda_9 \boldsymbol{\theta}_3) |\eta |^2|\rho |^2,  \nonumber \\
&\equiv & \textbf{h}^{\text{T}} \Lambda  \textbf{h}, 
\end{eqnarray}
where $\textbf{h}^{\text{T}}=(|\eta|^2,\, |\rho|^2,\,  |\chi|^2)\geq0$ and the matrix $\Lambda$ is 
\begin{eqnarray}
\label{eq:24}
\Lambda =
\begin{pmatrix}
\lambda_1 & (\lambda_6 +\lambda_9 \boldsymbol{\theta}_3)/2  & 
(\lambda_4 + \lambda_7 \boldsymbol{\theta}_1 + |\lambda_{10}| \boldsymbol{\theta}_4) /2\\
\star &\lambda_2 & (\lambda_5 +\lambda_8 \boldsymbol{\theta}_2)/2\\
\star & \star & \lambda_3
\end{pmatrix}.
\end{eqnarray}
Note that $V_4$ is a biquadratic form of the norm of the fields. Therefore, the scalar potential is bounded from below, $\textbf{h}^{\text{T}}\Lambda \textbf{h} > 0$, if the symmetric matrix  $\Lambda$ is strictly copositive, i.e. positive on  positive vectors $\textbf{h} > 0$. Note that we use the requirement of strong stability ($V_4>0$) because the stability in the marginal sense ($V_4\geq0$) does not allow any cubic terms in the scalar potential. However,  the term $\lambda_{15}\epsilon_{ijk}\eta_{i}\rho_{j}\chi_{k}$ must be included because in its absence appears a visible QCD axion (see Ref. \cite{sanchez2018new}). 

Mathematical conditions for a general symmetric matrix being strictly copositive were found in \cite{cottle1970classes,kaplan2000test} and applied in some elementary particle physics models in Refs. \cite{kannike2012vacuum, kannike2016vacuum}. For the case of a symmetric matrix A$_{3\times 3}$ of special interest for this paper,  these conditions read \cite{chang1994nonnegative, hadeler1983copositive}  
\begin{eqnarray}
\label{eq:25}
&&a_{11}>0, \quad a_{22}>0,\quad  a_{33}> 0,\nonumber\\
&&\bar{a}_{12}\equiv a_{12}+\sqrt{a_{11}a_{22}}>0, \quad \bar{a}_{13}\equiv a_{13}+\sqrt{a_{11}a_{33}}>0, \quad \bar{a}_{23}\equiv a_{23}+\sqrt{a_{22}a_{33}}>0,\nonumber\\
&&a_{12}\sqrt{a_{33}}+a_{13}\sqrt{a_{22}}+a_{23}\sqrt{a_{11}}+\sqrt{a_{11}a_{22}a_{33}}+\sqrt{2\bar{a}_{12}\bar{a}_{13}\bar{a}_{23}}> 0,
\end{eqnarray}
where $a_{ij}$ stands for a generic element of the A$_{3\times 3}$ matrix.

In order to apply these conditions in our particular case, it is convenient to rewrite the orbit parameters involved in the definition of $\Lambda$ as 
\begin{eqnarray}
\label{eq:26}
\boldsymbol{\theta}_1= \hat{\chi}^*_j\hat{\eta}_j\hat{\eta}^*_i\hat{\chi}_i=|\boldsymbol{\theta}_1|, 
\quad {\boldsymbol{\theta}}_2 =\hat{\chi}^*_j\hat{\rho}_j\hat{\rho}^*_i\hat{\chi}_i=|\boldsymbol{\theta}_2|, \nonumber \\
\quad {\boldsymbol{\theta}}_3= \hat{\eta}^*_i\hat{\rho}_i \hat{\rho}^*_j\hat{\eta}_j=|\boldsymbol{\theta}_3|,
\quad \boldsymbol{\theta}_4 = (\hat{\chi}^*_i \hat{\eta}_i)^2+ \mathrm{H.c.}=2|\boldsymbol{\theta}_1|\cos(\omega_{\boldsymbol{\theta}_1}),
\end{eqnarray}
where $\omega_{\boldsymbol{\theta}_1}$  is defined through $\hat{\chi}^*_j\hat{\eta}_j=\sqrt{|\boldsymbol{\theta}_1|}\exp(\omega_{\boldsymbol{\theta}_1}/2)$ and $0 \leq |\boldsymbol{\theta}_{1,2,3}|\leq 1$. These orbit parameters can be determined from the following observation. If $V_4<0$ for a particular direction with $|\textbf{h}| =1$ determined by a set of orbit parameters, then the scalar potential will not be bounded from below in the limit $|\textbf{h}| \longrightarrow\infty$.  Therefore, to determine whether $V_4$ is stable in the limit of large field values, it is sufficient to find its minimum on a $|\textbf{h}| =1$ hypersphere. Doing so, we immediately notice that $\omega_{\boldsymbol{\theta}_1}=\pi$. The remaining orbit parameters are calculated by noting that $V_4$ is a monotonic function of them. Thus, the minimum is in some $|\boldsymbol{\theta}_{i}|$ values (with $i=1,2,3$) on the frontier of their space. 

\begin{figure}[h!]
\centering
\includegraphics[scale=0.4]{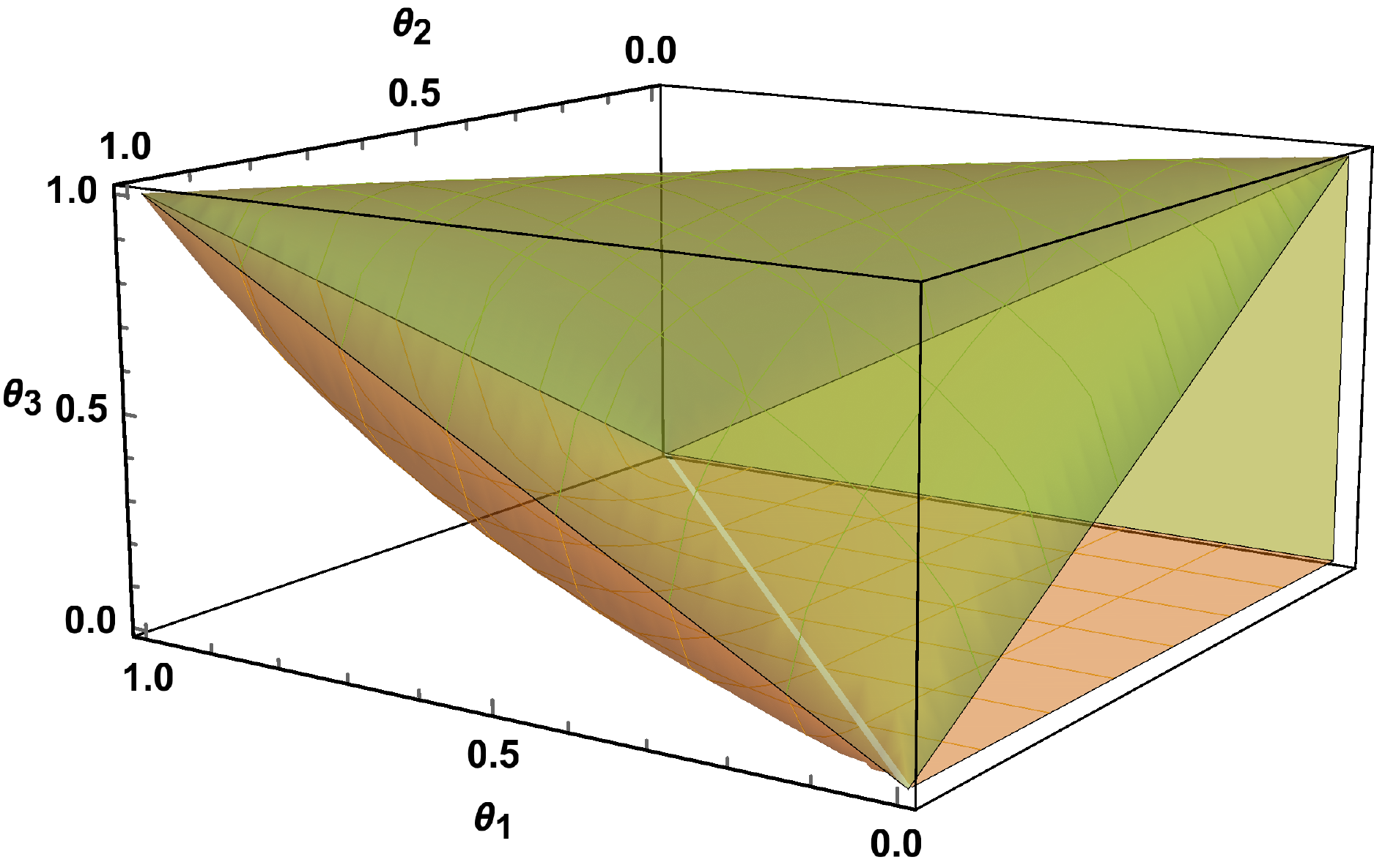}
\caption{Allowed space for the orbit parameters $\boldsymbol{\theta}_{1,2,3}$ defined in Eq. \eqref{eq:26}. Note that the allowed volume in this space is not a cube, but a closed volume with boundaries given in Eq. \eqref{eq:26a}. Here, it is important to remark that a similar reduction of the orbit parameter space happens in the context of the triplet seesaw model \cite{PhysRevD.92.075028}. In that case, the relation on the boundary between the two orbit parameters makes that the allowed space is not longer a box. Thus, the stability conditions at tree level in that model are modified to a set less restrictive, correcting what had been done in the previous literature.}
\label{frontier}
\end{figure}

At first glance, we could think the frontier for the $|\boldsymbol{\theta|}$ is a cube with edge length equals one. However, from the definition of $|\boldsymbol{\theta|}$ given in Eqs. \eqref{eq:26}, it can be found, without loss of generality, that the frontier is given by
\begin{eqnarray}
\label{eq:26a}
\resizebox{.3\hsize}{!}{$0 \leq |\boldsymbol{\theta}_{1}|\leq 1, \quad  0 \leq |\boldsymbol{\theta}_{2}|\leq 1,$} \nonumber \\
\resizebox{.9\hsize}{!}{$\max \left[0,\,\sqrt{|\boldsymbol{\theta}_{1}||\boldsymbol{\theta}_{2}|}-\sqrt{\left( 1-|\boldsymbol{\theta}_{1}|\right)\left( 1-|\boldsymbol{\theta}_{2}|\right)}  \right]^2 \leq |\boldsymbol{\theta}_{3}| \leq   \left[\sqrt{|\boldsymbol{\theta}_{1}||\boldsymbol{\theta}_{2}|}+\sqrt{\left( 1-|\boldsymbol{\theta}_{1}|\right)\left( 1-|\boldsymbol{\theta}_{2}|\right)}  \right]^2.$}  \nonumber \\
\end{eqnarray}
In Figure \ref{frontier} we can see that the space of $|\boldsymbol{\theta}_{1,2,3}|$ is actually smaller than the cube. In general, note that $|\boldsymbol{\theta}_{1,2,3}|$ are independent orbit parameters; however, their frontiers are fixed by Eq. \eqref{eq:26a}. 
Although we know the values of $|\boldsymbol{\theta}_{1,2,3}|$ which minimize $V_4$ on the hypersphere with $|\textbf{h}| =1$ are in the frontier given in Figure \ref{frontier}, their specific values depend on the $\lambda_i$ coefficients in the quartic scalar potential. Among the eight general cases, there are five cases where the minimizing $|\boldsymbol{\theta}_{1,2,3}|$ values can be easily found. These cases are:
\begin{itemize}
\item[(i)] $|\boldsymbol{\theta}_{1,2,3}|=0$, if $\lambda_7-2|\lambda_{10}|\geq0$, $\lambda_8\geq0$ and $\lambda_9\geq0$; 
\item[(ii)] $|\boldsymbol{\theta}_{1,2}|=0$ and $|\boldsymbol{\theta}_{3}|=1$, 
if $\lambda_7-2|\lambda_{10}|\geq0$, $\lambda_8\geq0$ and $\lambda_9<0$;  
\item[(iii)] $|\boldsymbol{\theta}_{1,3}|=0$ and $|\boldsymbol{\theta}_{2}|=1$, if
 $\lambda_7-2|\lambda_{10}|\geq0$, $\lambda_8<0$ and $\lambda_9\geq0$; 
\item[(iv)] $|\boldsymbol{\theta}_{2,3}|=0$ and $|\boldsymbol{\theta}_{1}|=1$, if
$\lambda_7-2|\lambda_{10}|<0$, $\lambda_8\geq0$ and $\lambda_9\geq0$;
\item[(v)] $|\boldsymbol{\theta}_{1,2,3}|=1$, if
$\lambda_7-2|\lambda_{10}|<0$, $\lambda_8<0$ and $\lambda_9<0$. 
\end{itemize} 
Now, by applying the copositivity conditions in Eq. \eqref{eq:25} to the symmetric matrix $\Lambda$ with the orbit parameters in the five previous cases, we find the following constraints on the $\lambda$ parameters of the scalar potential:  
\begin{eqnarray}
\label{eq:27}
& & \lambda_1 > 0,\quad \lambda_2 >0,\quad \lambda_3 >0,\nonumber\\
& &\lambda_4 +2\sqrt{\lambda_1 \lambda_3}>0, \quad \lambda_4 + \lambda_7  -2 |\lambda_{10}|+2\sqrt{\lambda_1 \lambda_3}>0, \nonumber \\
& &\lambda_5 + 2\sqrt{\lambda_2 \lambda_3}>0,\quad \lambda_5 +\lambda_8 + 2\sqrt{\lambda_2 \lambda_3}>0, \nonumber\\
& &\lambda_6 + 2\sqrt{\lambda_1 \lambda_2}>0, \quad 
\lambda_6 +\lambda_9 + 2\sqrt{\lambda_1 \lambda_2}>0, 
\end{eqnarray}
and 2$^3=8$ additional inequalities with the following form:  
\begin{eqnarray}
\label{eq:28}
\textrm{C}_{1}\sqrt{\lambda_2}+\textrm{C}_{2}\sqrt{\lambda_1}+\textrm{C}_{3}\sqrt{\lambda_3}+2\sqrt{\lambda_1 \lambda_2 \lambda_3}+\sqrt{\overline{\textrm{C}}_{1}\overline{\textrm{C}}_{2}\overline{\textrm{C}}_{3}}> 0,
\end{eqnarray} 
where $ \textrm{C}_1=\lbrace \lambda_4,\,\lambda_4 + \lambda_7 -2 |\lambda_{10}|\rbrace$, $ \textrm{C}_2=\lbrace  \lambda_5,\,\lambda_5 +\lambda_8 \rbrace$, $ \textrm{C}_3= \lbrace  \lambda_6,\,\lambda_6 +\lambda_9 \rbrace$, $\overline{\textrm{C}}_1= \textrm{C}_1+2\sqrt{\lambda_1\lambda_{3}}$, $ \overline{\textrm{C}}_2=\textrm{C}_2+2\sqrt{\lambda_2\lambda_{3}}$ and $\overline{\textrm{C}}_3=\textrm{C}_3+2\sqrt{\lambda_1\lambda_{2}}$. Note that these 17 inequalities are the same as those obtained if we had considered the cube with side one as the orbit parameter space. 

The remaining three cases: (vi) $\lambda_7-2|\lambda_{10}|\geq0$, $\lambda_8<0$ and $\lambda_9<0$;  (vii) $\lambda_7-2|\lambda_{10}|<0$, $\lambda_8\geq0$ and $\lambda_9<0$;  and (viii) $\lambda_7-2|\lambda_{10}|<0$, $\lambda_8<0$ and $\lambda_9\geq0$ are more complicated because the minimizing $|\boldsymbol{\theta}_{1,2,3}|$ parameters, $|\boldsymbol{\theta}_{1,2,3}|^{\star}$, depend  not only on the signal of  $\lambda_7-2|\lambda_{10}|$, $\lambda_8$ and $\lambda_9$ but also on the values of the $\lambda_i$ couplings . In other words, the $|\boldsymbol{\theta}_{1,2,3}|^{\star}$  are now functions of $\lambda_i$ with values on the frontier in Figure \ref{frontier}. In general, these cases bring inequalities that can be written as (see for example \cite{PhysRevD.92.075028}):  $\quad \lambda_4 + (\lambda_7  -2 |\lambda_{10}|)|\boldsymbol{\theta}_{1}|^{\star}+2\sqrt{\lambda_1 \lambda_3}>0$, $\lambda_5 +\lambda_8\,|\boldsymbol{\theta}_{2}|^{\star}  + 2\sqrt{\lambda_2 \lambda_3}>0$ and $\lambda_6 +\lambda_9\,|\boldsymbol{\theta}_{3}|^{\star}  + 2\sqrt{\lambda_1 \lambda_2}>0$. Although  $|\boldsymbol{\theta}_{1,2,3}|^{\star}(\lambda_i)$ depend on  the specific values of $\lambda_i$ couplings, these still satisfy that $0\leq|\boldsymbol{\theta}_{1,2,3}|^{\star}(\lambda_i)\leq 1$. That is a key point, because of that, we do not really need to calculate these intricate functions. In order to clarify that point, let's consider the particular case of $\lambda_8<0$. As $0\leq |\boldsymbol{\theta}_{2}|^{\star}(\lambda_i)\leq 1$ and $\lambda_8<0$, then $0 \geq \lambda_8|\boldsymbol{\theta}_{2}|^{\star}(\lambda_i) \geq \lambda_8$. Applying general properties of inequalities, we find $2\sqrt{\lambda_2\lambda_3} \geq \lambda_8|\boldsymbol{\theta}_{2}|^{\star}(\lambda_i)+2\sqrt{\lambda_2\lambda_3}  \geq \lambda_8+2\sqrt{\lambda_2\lambda_3}$ or in an equivalent way, 
\begin{equation}
\label{eq:28a}
 -\lambda_8-2\sqrt{\lambda_2\lambda_3}\geq -\lambda_8|\boldsymbol{\theta}_{2}|^{\star}(\lambda_i)-2\sqrt{\lambda_2\lambda_3} \geq -2\sqrt{\lambda_2\lambda_3}. 
\end{equation}
Now,  from $\lambda_5 +\lambda_8 + 2\sqrt{\lambda_2 \lambda_3}>0$ inequality in Eq. \eqref{eq:28}, we find $\lambda_5>-\lambda_8 - 2\sqrt{\lambda_2 \lambda_3}\geq -\lambda_8|\boldsymbol{\theta}_{2}|^{\star}(\lambda_i)-2\sqrt{\lambda_2\lambda_3}$ where we have used Eq. \eqref{eq:28a} in the last step. Now, we realize that $\lambda_5>\lambda_8+2\sqrt{\lambda_2 \lambda_3}\geq \lambda_8|\boldsymbol{\theta}_{2}|^{\star}(\lambda_i)+2\sqrt{\lambda_2\lambda_3}$ implies that if we satisfy the $\lambda_5 +\lambda_8 + 2\sqrt{\lambda_2 \lambda_3}>0$ inequality, the unknown $\lambda_5 +\lambda_8\,|\boldsymbol{\theta}_{2}|^{\star} +2\sqrt{\lambda_2 \lambda_3}>0$ inequality is also satisfied when $\lambda_8<0$. We remark here that this result is independent on the particular form of  $|\boldsymbol{\theta}_{2}|^{\star}$ provided $0\leq|\boldsymbol{\theta}_{2}|^{\star}(\lambda_i)\leq 1$ is valid. For the case of $\lambda_8>0$, we can follow an analogue procedure to arrive to the conclusion that for this case when the $\lambda_5 + 2\sqrt{\lambda_2 \lambda_3}>0$ inequality in Eq. \eqref{eq:28} is satisfied, then  $\lambda_5 +\lambda_8\,|\boldsymbol{\theta}_{2}|^{\star} +2\sqrt{\lambda_2 \lambda_3}>0$ is automatically satisfied. Therefore, whatever the sign of $\lambda_8$ is, the unknown $\lambda_5 +\lambda_8\,|\boldsymbol{\theta}_{2}|^{\star} +2\sqrt{\lambda_2 \lambda_3}>0$ is not necessary. Similar conclusions are obtained for the remaining two unknown inequalities: $\quad \lambda_4 + (\lambda_7  -2 |\lambda_{10}|)|\boldsymbol{\theta}_{1}|^{\star}+2\sqrt{\lambda_1 \lambda_3}>0$ and $\lambda_6 +\lambda_9\,|\boldsymbol{\theta}_{3}|^{\star}  + 2\sqrt{\lambda_1 \lambda_2}>0$. Therefore, we have only 17 necessary and sufficient inequalities given by Eqs. \eqref{eq:27} and \eqref{eq:28} coming from the strict copositivity of $\Lambda$, which additionally delimit the scalar potential parameters. 

\begin{figure}[h!]
\centering
\includegraphics[scale=0.6]{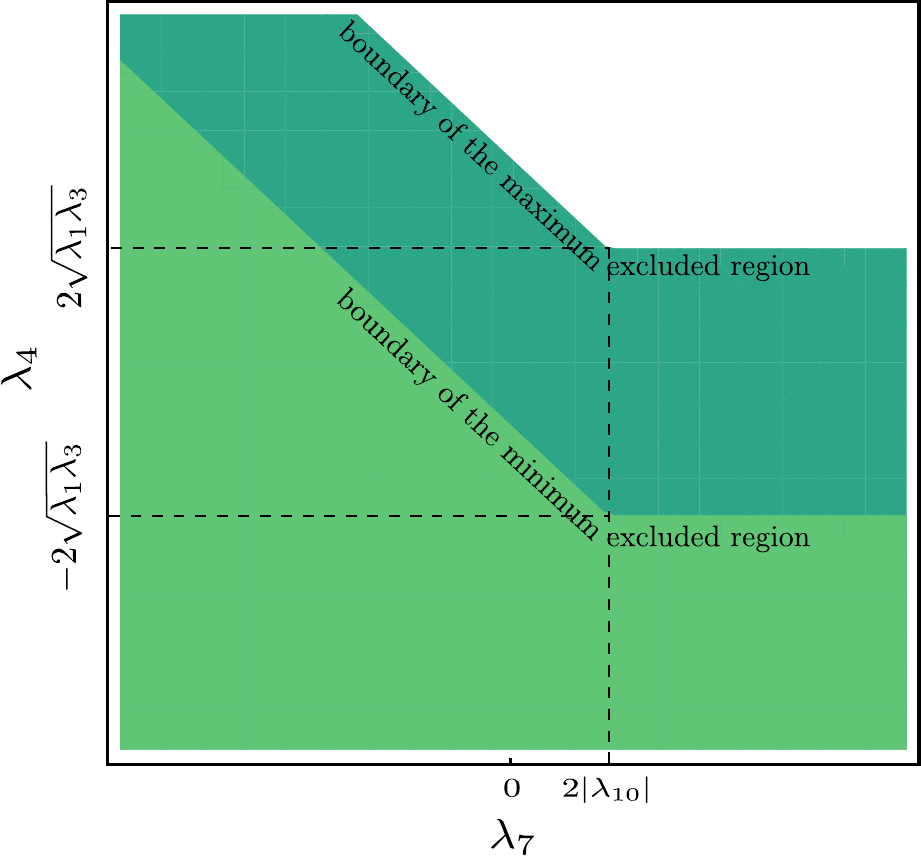}
\caption{The maximum excluded region for $\lambda_4$ is defined by $\lambda_4\leq2\sqrt{\lambda_1\lambda_3}$ for $\lambda_7\geq2|\lambda_{10}|$, and $\lambda_4\leq2\sqrt{\lambda_1\lambda_3}+2|\lambda_{10}|-\lambda_7$ for $\lambda_7<2|\lambda_{10}|$, whereas the minimum excluded region is defined by $\lambda_4\leq-2\sqrt{\lambda_1\lambda_3}$ for $\lambda_7\geq2|\lambda_{10}|$, and $\lambda_4\leq-2\sqrt{\lambda_1\lambda_3}+2|\lambda_{10}|-\lambda_7$ for $\lambda_7<2|\lambda_{10}|$.}
\label{maxminreg4}
\end{figure}

\begin{figure}[h!]
\centering
\begin{subfigure}{.5\textwidth}
  \centering
  \includegraphics[width=.8\linewidth]{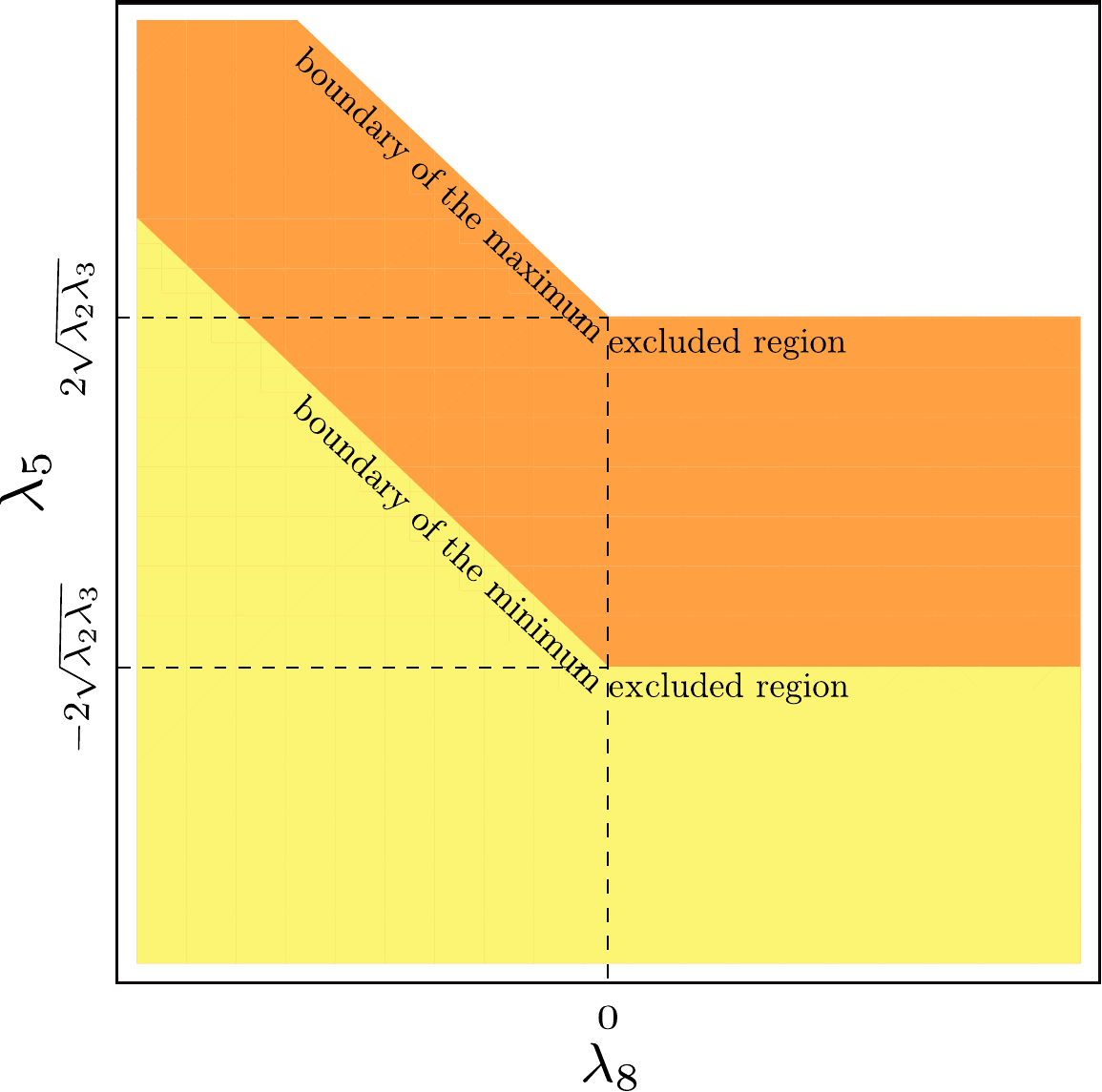}
\end{subfigure}%
\begin{subfigure}{.5\textwidth}
  \centering
  \includegraphics[width=.83\linewidth]{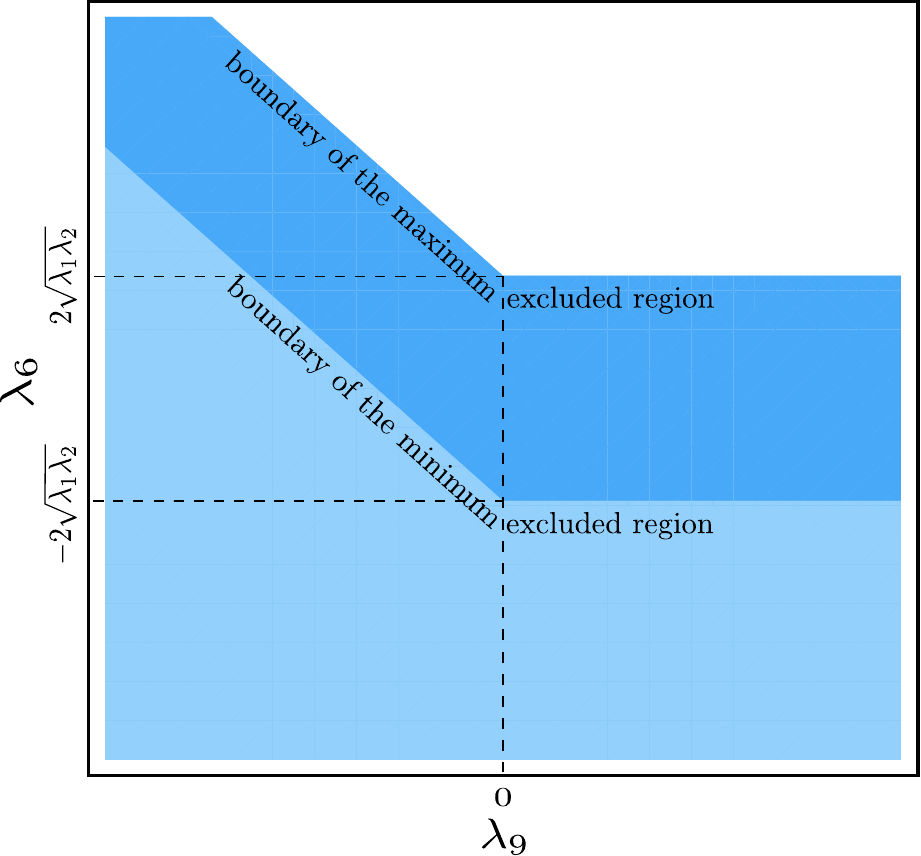}
\end{subfigure}
\caption{Maximum and minimum excluded regions in the $\lambda_5-\lambda_8$ (left) and $\lambda_6-\lambda_9$ (right) parameter space. In general, the maximum and minimum excluded regions for $\lambda_5$ ($\lambda_6$) depend on the sign of $\lambda_8$ ($\lambda_9$) as explained in the text.}
\label{L56}
\end{figure}

At first glance, obtaining analytical results from \eqref{eq:27} and \eqref{eq:28} seems a very complicated task. Nevertheless, there are general relations between some $\lambda$'s that we can find. First, we can answer, for instance, the question: which is the largest excluded region in the $\lambda_4-\lambda_7$ plane provided the rest of the $\lambda$ couplings satisfy the copositivity conditions? That excluded region depends on the value of $\lambda_7$. On the one hand, if it is bigger than or equal to $2|\lambda_{10}|$, all $\lambda_4$ values smaller than or equal to $2\sqrt{\lambda_1\lambda_3}$ are excluded. On the other hand, for values of $\lambda_7$ smaller than $2|\lambda_{10}|$, the region $\lambda_4\leq2\sqrt{\lambda_1 \lambda_3}+2|\lambda_{10}|-\lambda_{7}$ is rejected. In both cases of $\lambda_7$, the largest excluded region occur whenever $\lambda_5=\lbrace-2\sqrt{\lambda_2\lambda_3},-2\sqrt{\lambda_2\lambda_3}-\lambda_8\rbrace$ and $\lambda_6=\lbrace-2\sqrt{\lambda_1\lambda_2},-2\sqrt{\lambda_1\lambda_2}-\lambda_9\rbrace$ where the first value of $\lambda_5\ (\lambda_6)$ corresponds to $\lambda_8\geq0\ (\lambda_9\geq0)$ and the second one corresponds to $\lambda_8<0(\lambda_9<0)$, independently of the other values of $\lambda$ couplings. Once $\lambda_5$ and $\lambda_6$ take different values from the aforementioned ones, the excluded region for $\lambda_4$ decreases. However, that region does not decrease forever since there is also a lower bound depending on 
$\lambda_7$ coupling. In more detail, if $\lambda_7$ is bigger than or equal to $2|\lambda_{10}|$, all $\lambda_4$ values smaller than or equal to $-2\sqrt{\lambda_1\lambda_3}$ are always excluded. For the case of $\lambda_7$ values smaller than or equal to $2|\lambda_{10}|$, $\lambda_4$ values smaller than or equal to $-2\sqrt{\lambda_1\lambda_3}+2|\lambda_{10}|-\lambda_7$ are always excluded.
For both cases, points satisfying the following conditions are on the bound:
\begin{itemize}
\item[(i)] $\lambda_5\sqrt{\lambda_1}+\lambda_6\sqrt{\lambda_3}>0$, if 
$\lambda_8\geq 0$ and 
$\lambda_9\geq 0$.
\item[(ii)] 
$\lambda_5\sqrt{\lambda_1}+(\lambda_6+\lambda_9)\sqrt{\lambda_3}>0$, if 
$\lambda_8\geq 0$ and 
$\lambda_9< 0$.
\item[(iii)] $(\lambda_5+\lambda_8)\sqrt{\lambda_1}+\lambda_6\sqrt{\lambda_3}>0$, if $\lambda_8<0$ and $\lambda_9\geq0$.
\item[(iv)] $(\lambda_5+\lambda_8)\sqrt{\lambda_1}+(\lambda_6+\lambda_9)\sqrt{\lambda_3}>0$, if $\lambda_8<0$ and $\lambda_9<0$.
\end{itemize} 
The maximum and minimum excluded regions for $\lambda_4$ as a function of $\lambda_7$ are shown in Fig. \ref{maxminreg4}. Similar conclusions can be reached for the $\lambda_5-\lambda_8$ and $\lambda_6-\lambda_9$ planes and are shown in Fig. \ref{L56}. In the left panel of Fig. \ref{L56}, the maximum and the minimum excluded regions for $\lambda_5$ as a function of $\lambda_8$ are shown. In general, the maximum excluded region of $\lambda_5$ depends on the sign of $\lambda_8$, i.e. if it is negative, then $\lambda_5$ values satisfying $\lambda_5\leq2\sqrt{\lambda_2\lambda_3}-\lambda_8$ are excluded. Otherwise, $\lambda_5$ values smaller than or equal to $2\sqrt{\lambda_2\lambda_3}$ are excluded.  The maximum excluded region of $\lambda_5$ is reached when 
$\lambda_4 + \left( \lambda_7 -2 |\lambda_{10}|\right)\theta_\textrm{H}\left(2|\lambda_{10}| -\lambda_7\right)+2\sqrt{\lambda_1\lambda_{3}}=0$ and $\lambda_6 +\lambda_9\,\theta_\textrm{H}\left(-\lambda_9\right) +2\sqrt{\lambda_1\lambda_{2}}=0$, where $\theta_\textrm{H}$ is the Heaviside theta function.  The minimum excluded region of $\lambda_5$ has the same dependence on $\lambda_8$ as the maximum excluded one, however, its bounds are different. If $\lambda_8$ is smaller than zero, then $\lambda_5$ values satisfying $\lambda_5\leq-2\sqrt{\lambda_2\lambda_3}-\lambda_8$ are excluded. Otherwise, $\lambda_5\leq-2\sqrt{\lambda_2\lambda_3}$ are excluded. These bounds are attained when $\left(\lambda_4 + \left( \lambda_7 -2 |\lambda_{10}|\right)\theta_\textrm{H}\left(2|\lambda_{10}| -\lambda_7\right)+2\sqrt{\lambda_1\lambda_{3}}\right)\,\sqrt{\lambda_2}+\left( \lambda_6 +\lambda_9\,\theta_\textrm{H}\left(-\lambda_9\right) +2\sqrt{\lambda_1\lambda_{2}}\right)\,\sqrt{\lambda_3}>4\sqrt{\lambda_1\lambda_2\lambda_3}$. 

On the other hand, in the right panel of Fig. \ref{L56}, the maximum and the minimum excluded regions for $\lambda_6$ as a function of $\lambda_9$ are shown. The maximum excluded region of $\lambda_6$ is characterized by two different bounds. $\lambda_6\leq2\sqrt{\lambda_1\lambda_2}-\lambda_9$ are excluded if $\lambda_9<0$ and $\lambda_6$ values satisfying $\lambda_6\leq2\sqrt{\lambda_1\lambda_2}$ are excluded if $\lambda_9\geq0$. That region is reached when $\lambda_4 + \left( \lambda_7 -2 |\lambda_{10}|\right)\theta_\textrm{H}\left(2|\lambda_{10}| -\lambda_7\right)$ $+2\sqrt{\lambda_1\lambda_{3}}=0$
and $\lambda_5 +\lambda_8\,\theta_\textrm{H}\left(-\lambda_8\right)+2\sqrt{\lambda_2\lambda_{3}}=0$. Finally, the minimum excluded region also has two different bounds. $\lambda_6\leq-2\sqrt{\lambda_1\lambda_2}-\lambda_9$ are excluded if $\lambda_9<0$ and $\lambda_6$ values satisfying $\lambda_6\leq-2\sqrt{\lambda_1\lambda_2}$ are excluded if $\lambda_9\geq0$. That region is reached when $\left(\lambda_4 + \left( \lambda_7 -2 |\lambda_{10}|\right)\theta_\textrm{H}\left(2|\lambda_{10}| -\lambda_7\right)+2\sqrt{\lambda_1\lambda_{3}}\right)\sqrt{\lambda_2}$  \newline
$+\left(\lambda_5 +\lambda_8\,\theta_\textrm{H}\left(-\lambda_8\right)\right.$ $ \left. 
+2\sqrt{\lambda_2\lambda_{3}}\right)\sqrt{\lambda_1}$ $ >4\sqrt{\lambda_1\lambda_2\lambda_3}$.

 
\begin{figure}[h!]
\centering
\includegraphics[scale=0.2]{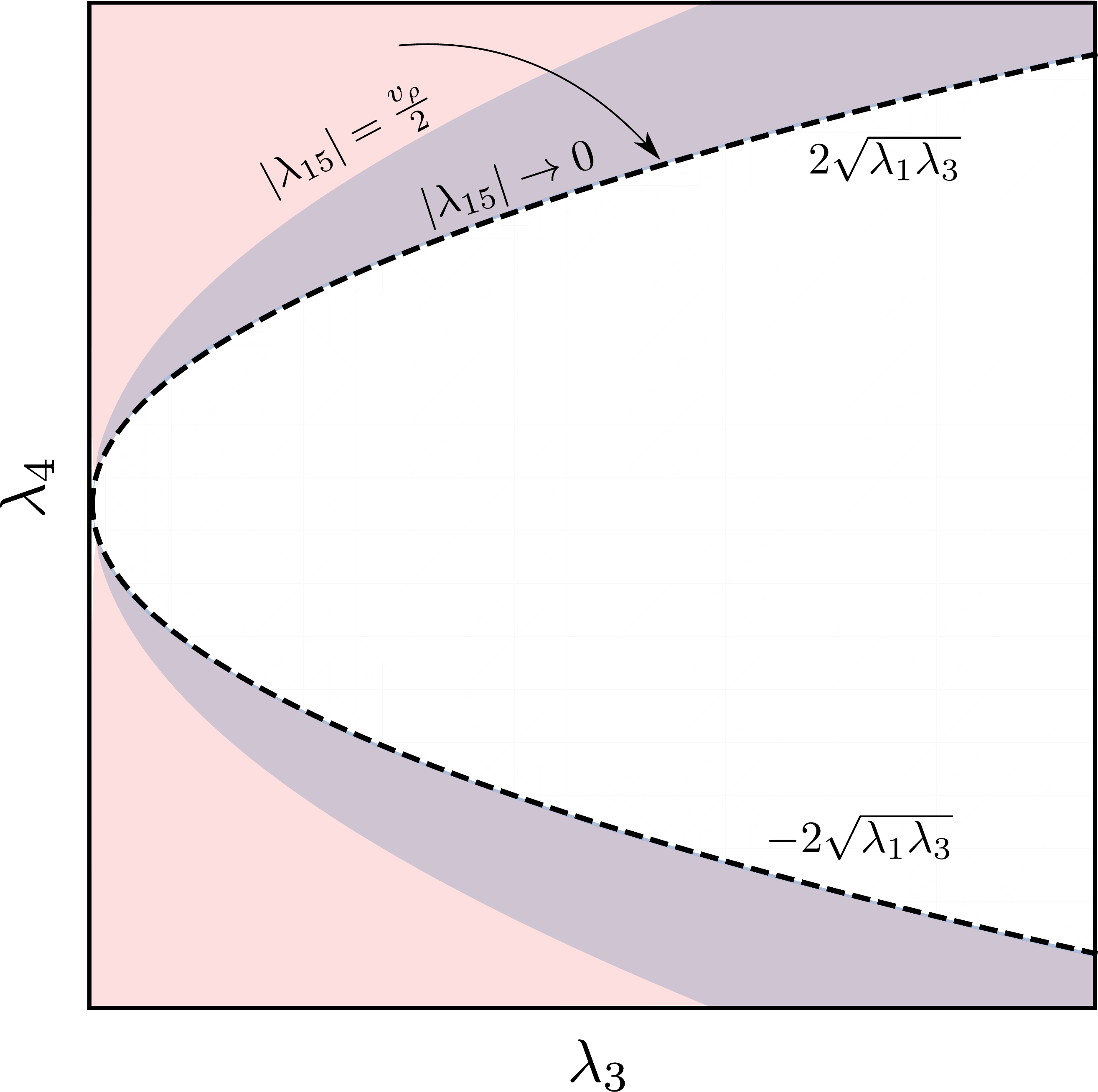}
\caption{The excluded region for $\lambda_4$ coming from Hessian matrix criterion increases when $|\lambda_{15}|$ decreases, and it also approaches to excluded region given by copositiviy. Similar conclusions can be reached for $\lambda_5$ and $\lambda_6$ couplings.}
\label{L4uplim}
\end{figure}

It is also interesting to combine the copositivity conditions in Eqs. \eqref{eq:27} and \eqref{eq:28} with those coming from the positivity of the Hessian matrix in Eq. \eqref{eq:22}. Doing that, it is straightforward to see that the three conditions in the first line in Eq. \eqref{eq:22} are always satisfied provided $\lambda_1 > 0,\, \lambda_2 >0,\, \lambda_3 > 0$ as required by copositivity. In addition, the lower bounds on $\lambda_{4,5,6}$ coming from Eq. \eqref{eq:22} are always smaller than the obtained from copositivity. More specifically, the lower bound on $\lambda_{4}$, for instance, coming from the Hessian matrix can be written as $\lambda_4>\,-2\sqrt{\lambda_1 \lambda_3} - \frac{|\lambda_{15}|}{4v_\rho}\left( \frac{v_\rho^2}{v_\eta v_\chi} \right) \left( \sqrt{\frac{\lambda_3}{\lambda_1}} \frac{v_\chi^2}{v_\eta^2} + \sqrt{\frac{\lambda_1}{\lambda_3}} \frac{v_\eta^2}{v_\chi^2} -2  \right)$, where we have assumed that $|\lambda_{15}| \ll v_\eta,v_\rho$. As $\left( \sqrt{\frac{\lambda_3}{\lambda_1}} \frac{v_\chi^2}{v_\eta^2} + \sqrt{\frac{\lambda_1}{\lambda_3}} \frac{v_\eta^2}{v_\chi^2} -2  \right)$ is always positive, we conclude that the lower bound obtained from copositivity is stronger than the Hessian one. It is also important  to note that the upper limit on $\lambda_{4,5,6}$ coming from the positivity of the Hessian matrix becomes stronger when $|\lambda_{15}| \ll v_\eta,v_\rho$. For instance, in the limit $|\lambda_{15}|\rightarrow0$, these limits go to $\lambda_4<2\sqrt{\lambda_1\lambda_3}$, $\lambda_5<2\sqrt{\lambda_2\lambda_3}$ and $\lambda_6<2\sqrt{\lambda_1\lambda_2}$ which are the bounds on $\lambda_{4,5,6}$ of the maximum excluded region discussed previously. For more details on the behaviour of the upper and lower bounds of $\lambda_4$ as a function of $|\lambda_{15}|$ see Fig. \ref{L4uplim}.  Similar conclusions are obtained for $\lambda_5$ and $\lambda_6$ couplings. It is important to note that since the term $\frac{\lambda_{15}}{\sqrt{2}}\epsilon_{ijk}\eta_{i}\rho_{j}\chi_{k}$ softly breaks the $\ZZ_2$ symmetry, it is technically natural that $\lambda_{15}$ gets a small value in comparison with other energy levels in the model, $v=\sqrt{v_\eta^2+v_\rho^2}$ and $v_\chi$, since making $\lambda_{15}\rightarrow0$ increases the symmetry of the total Lagrangian.

On the other hand,  copositivity does not impose upper limits on $\lambda_{4,5,6}$ and therefore, limits from Eq. \eqref{eq:22} still apply, i.e. 
\begin{eqnarray}
\label{eq:29}
\textrm{bounds from copositivity}< &\lambda_4 & < \frac{ |\lambda_{15}| v_\rho +\sqrt{(H_0)_{11}(H_0)_{33}}}{2v_\eta v_\chi}, \nonumber \\
\textrm{bounds from copositivity} < &\lambda_5 & < \frac{ |\lambda_{15}|v_\eta +\sqrt{(H_0)_{22}(H_0)_{33}}}{2 v_\rho v_\chi}  ,\nonumber \\ 
\textrm{bounds from copositivity}<&\lambda_6 & < \frac{|\lambda_{15}| v_\chi +\sqrt{(H_0)_{11}(H_0)_{22}}}{2 v_\eta v_\rho}, 
\end{eqnarray}
where $(H_0)_{ii}$ with $i=1,2,3$ is the corresponding element of the Hessian matrix. 
In order to conclude this section, we can say that out of fourteen initial parameters in the scalar potential in Eq. \eqref{eq:16}, we can clearly constrain ten of them by applying the copositivity criterion in addition to the first and second derivative tests. Specifically, the $\mu_{1,2,3}$ parameters in Eq. \eqref{eq:20}, the $0\leq\lambda_{1,2,3,10}\leq4\pi$ and $\lambda_{4,5,6}$ in Eq. \eqref{eq:29} are explicitly constrained. Finally, needless to say that apart from the $\lambda_{15}$ (which has dimension of mass), the rest of the $\lambda$ parameters are constrained by perturbativity to be $|\lambda|\leq4\pi$.


\section{Scalar mass spectrum\label{scalarmasses}}

A complementary set of constraints on the $\lambda$'s comes from the positivity of the square masses of the scalars of this model. To be more specific, the physical spectrum of scalars consists of four charged scalars, $H^{\pm}$ and $H_{V}^{\pm},$ one complex neutral field, $H_U$, one CP-odd scalar, $H_0$, and three CP-even scalars, $h$, $H_2$ and $H_3$, from which we define $h$ as the SM Higgs boson.  Their square masses can be written as:
\begin{eqnarray}
&& M_{H_0}^2=\frac{|\lambda_{15}|}{2v_{\eta}v_{\rho}v_{\chi}}(v_{\eta}^2v_{\rho}^2+(v_{\eta}^2+v_{\rho}^2)v_{\chi}^2), \quad M_{H_U}^2=\frac{v_{\eta}^2+v_{\chi}^2}{2v_{\eta}v_{\chi}}(|\lambda_{15}|v_{\rho}+\lambda_7v_{\eta}v_{\chi}), 
\label{eq:30}\\
&& M_{H_{V}^{\pm}}^2=\frac{v_{\rho}^2+v_{\chi}^2}{2v_{\rho}v_{\chi}}(|\lambda_{15}|v_{\eta}+\lambda_8v_{\rho}v_{\chi}), \quad M_{H^{\pm}}^2=\frac{v^2}{2v_{\eta}v_{\rho}}(|\lambda_{15}|v_{\chi}+\lambda_9v_{\eta}v_{\rho}). \label{eq:31}
\end{eqnarray}
At this point, it is important to recall that $\lambda_{15}$ has dimension of mass.  We have not yet included the CP-even scalars in Eqs. (\ref{eq:30}-\ref{eq:31}) because their masses are more intricate and require a detailed analysis, as we show below.
 
From Eq. \eqref{eq:30} we note that $ M_{H_0}^2$ is always larger than or equal to zero. However, the positivity of $M_{H_U}^2$ requires that
\begin{eqnarray}
\label{eq:32}
&& \lambda_7 \geq -\frac{v_{\rho}}{v_{\eta}v_{\chi}}|\lambda_{15}|.
\end{eqnarray}
For the case of charged scalars a stronger limit can be applied because their square masses not only have to be positive but also $\gtrsim 80^2$ GeV$^2$ ($95\%$  C.L.) \cite{tanabashi2018aps}. Applying this constraint to $M_{H^{\pm}}^2$ and $M_{H_{V}^{\pm}}^2$, we obtain 
\begin{eqnarray}
\label{eq:33}
&& \lambda_{8}\geq \frac{2\times 80^2}{(v_{\rho}/\textrm{GeV})^2+(v_{\chi}/\textrm{GeV})^2}-\frac{v_{\eta}}{v_{\rho}v_{\chi}}|\lambda_{15}|,\quad \textrm{and} \quad
\lambda_{9}\geq 2\times\frac{80^2}{246^2}-\frac{v_{\chi}}{v_{\eta}v_{\rho}}|\lambda_{15}|.\nonumber \\
\end{eqnarray}

\begin{figure}[h!]
\centering
\includegraphics[scale=0.2]{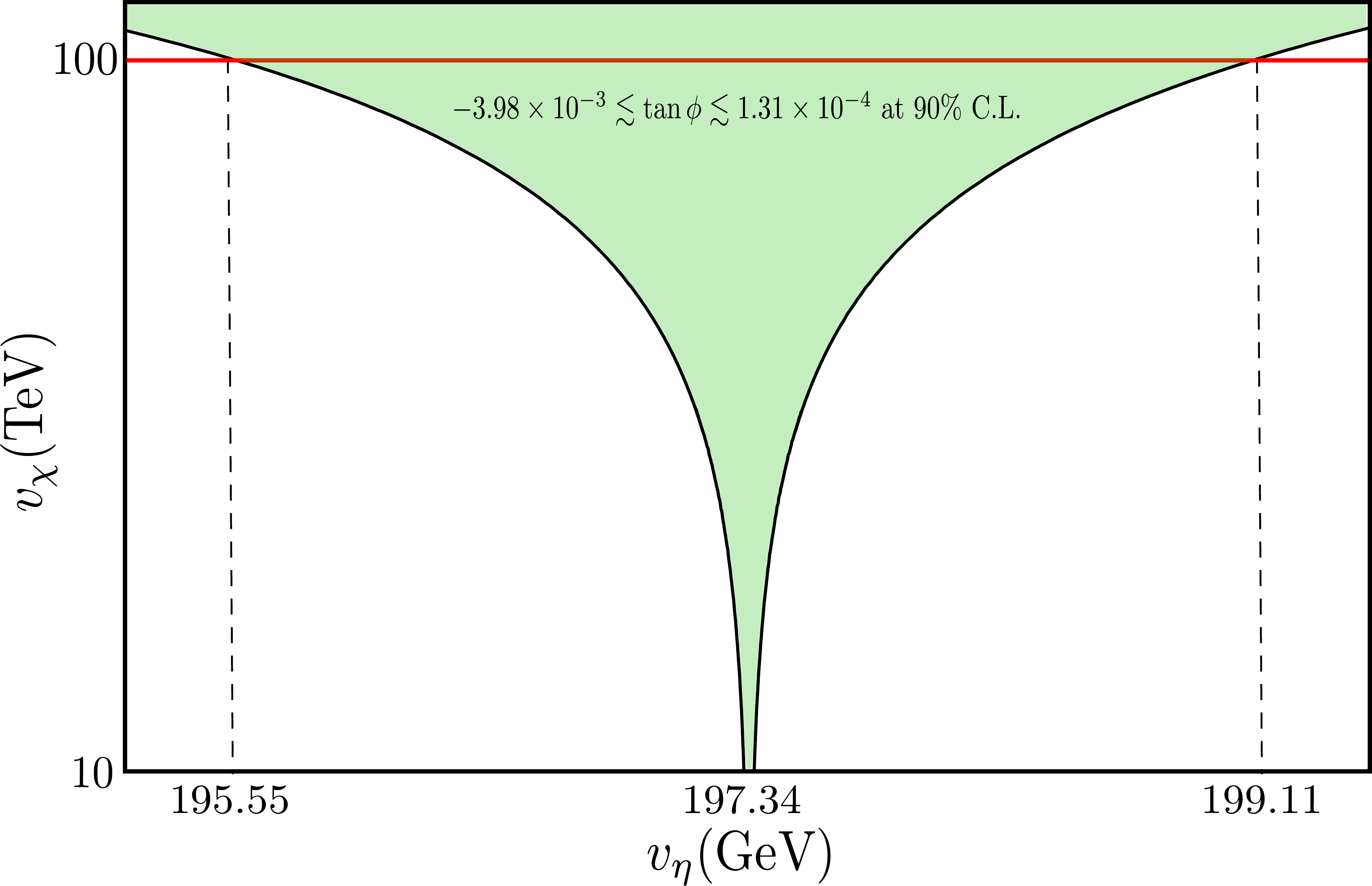}
\caption{The light green region contains the values for $v_\eta$ and $v_\chi$ allowing $-3.98\times$10$^{-3}\lesssim$ $\tan\phi \lesssim 1.31\times 10^{-4}$. $v_\eta$ values at least $1\%$ different from $197.34$ GeV require $v_\chi \gtrsim100$ TeV.}
\label{tangente}
\end{figure}

In order to estimate the lower limits for $\lambda_7, \, \lambda_8$ and $\lambda_9$ we bring in another piece of information. In general, in models with an extra neutral gauge boson, $Z_2$, the mixing angle, $\phi$, between this and the neutral gauge boson of the SM, $Z_1$,  has to be very small. For this $3-3-1$ model, lower and upper bounds  for this quantity have been found through the analysis of the $Z_1$ invisible- and charged lepton- partial decay width plus theoretical consistency of the $3-3-1$ models regarding the number of lepton families to be exactly 3. As a result, $-3.98\times$10$^{-3}\lesssim\tan\phi \lesssim 1.31\times 10^{-4}$ at 90$\%$ C.L. \citep{cogollo2008fermion}, independently of $M_{Z_2}$ and the hadronic sector. Other studies regarding FCNC suppression and parity violation in the Cesium atom have shown similar results \cite{long2001atomic,carcamo2006z}. As $\tan\phi$ depends on $v_{\eta}$ and $v_{\chi}$, we can estimate their values in order to satisfy this limit. To do so, let's write $\tan ^2\phi$ as \cite{van2005u}
\begin{equation}
\label{eq:34}
\tan^2\phi=\frac{M_Z^2-M_{Z_1}^2}{M_{Z_2}^2-M_Z^2},
\end{equation}
where the masses of the neutral gauge bosons $Z_1$ and $Z_2$ read
\begin{eqnarray}
\label{eq:35}
M_{Z_1}^2,\,M_{Z_2}^2&=&\frac{1}{2}\left[M_Z^2+M_{Z'}^2\mp\sqrt{(M_Z^2-M_{Z'}^2)^2+(2M_{ZZ'}^2)^2}\right],
\end{eqnarray} 
with
\begin{eqnarray}
\label{eq:36}
&&M_Z^2=\frac{g^2}{4\cos^2\theta_W}\,v^2,\nonumber\\
&&M_{Z'}^2=\frac{g^2}{12(1-\frac{1}{3}\tan^2\theta_W)}[(1+\tan^2\theta_W)^2 v^2-4\tan^2\theta_W v_{\eta}^2+4v_{\chi}^2],\nonumber\\
&&M_{ZZ'}^2=-\frac{g^2}{4\cos^2\theta_W}\frac{v^2-2\cos^2 \theta_W v_{\eta}^2}{\sqrt{3-4\sin^2\theta_W}}.
\end{eqnarray}
where $\theta_W$ is the Weinberg's angle and $g$ is the coupling constant of the SU$(2)_L$ group. Using $v=246$ GeV, $g=0.65$, $\sin^2 \theta_W\simeq 0.22 $,  $M_{Z_1}\simeq 91.19$ GeV \cite{ALEPH:2005ab,tanabashi2018aps} and imposing $-3.98\times$10$^{-3}\lesssim\tan\phi \lesssim 1.31\times 10^{-4}$ we find the allowed region in the $v_\chi-v_\eta$ plane. As we can see in Fig. \ref{tangente}, $v_\chi \gtrsim100$ TeV is needed to obtain significant deviations from the zero-mixing value, $v_\eta \simeq 197.34$ GeV. 

From the phenomenological point of view, $v_\chi \gtrsim 100 \textrm{ TeV}$ is not interesting due to the fact that all new particles in the model have masses of the order of $v_\chi$ and therefore would be unattainable by the current particle colliders. Thus, we use  $v_\eta \simeq 197.34$ GeV and $v_\rho = \sqrt{v^2 - v^2_{\eta}} \simeq 146.88$ GeV. Now, it is also possible to estimate $v_\chi$ from the fact that the mass of an extra neutral gauge boson, $Z_2$, has to be larger than $2.9$ TeV \cite{Sirunyan:2017nrt,tanabashi2018aps}. Applying this bound in Eq. \eqref{eq:35} and using the values for $v_\eta$ and $v_\rho$ obtained above, we have that $v_\chi \gtrsim 7.31$ TeV. Thus, we can use these VEVs to estimate lower limits for $\lambda_7, \, \lambda_8$ and $\lambda_9$ in terms of the value of $|\lambda_{15}|$
\begin{eqnarray}
\label{eq:37}
&&\lambda_7\gtrsim-1.02 \, \frac{|\lambda_{15}|}{\textrm{GeV}}, \,\,   \lambda_{8}\gtrsim 0.02-1.8\times10^{-4}\,\frac{|\lambda_{15}|}{\textrm{GeV}},\,\,  \lambda_{9}\gtrsim 0.21-0.25\,\frac{|\lambda_{15}|}{\textrm{GeV}},
\end{eqnarray}
where we have used $v_\chi= 7.31$ TeV. In the case that $\lambda_{15}\ll 1$ GeV, for instance, we have  $\lambda_7\gtrsim 0, \, \lambda_8\gtrsim 0.02$ and $\lambda_9\gtrsim 0.21$.

\begin{figure}[h!]
\centering
\begin{subfigure}{.5\textwidth}
  \centering
  \includegraphics[width=.8\linewidth]{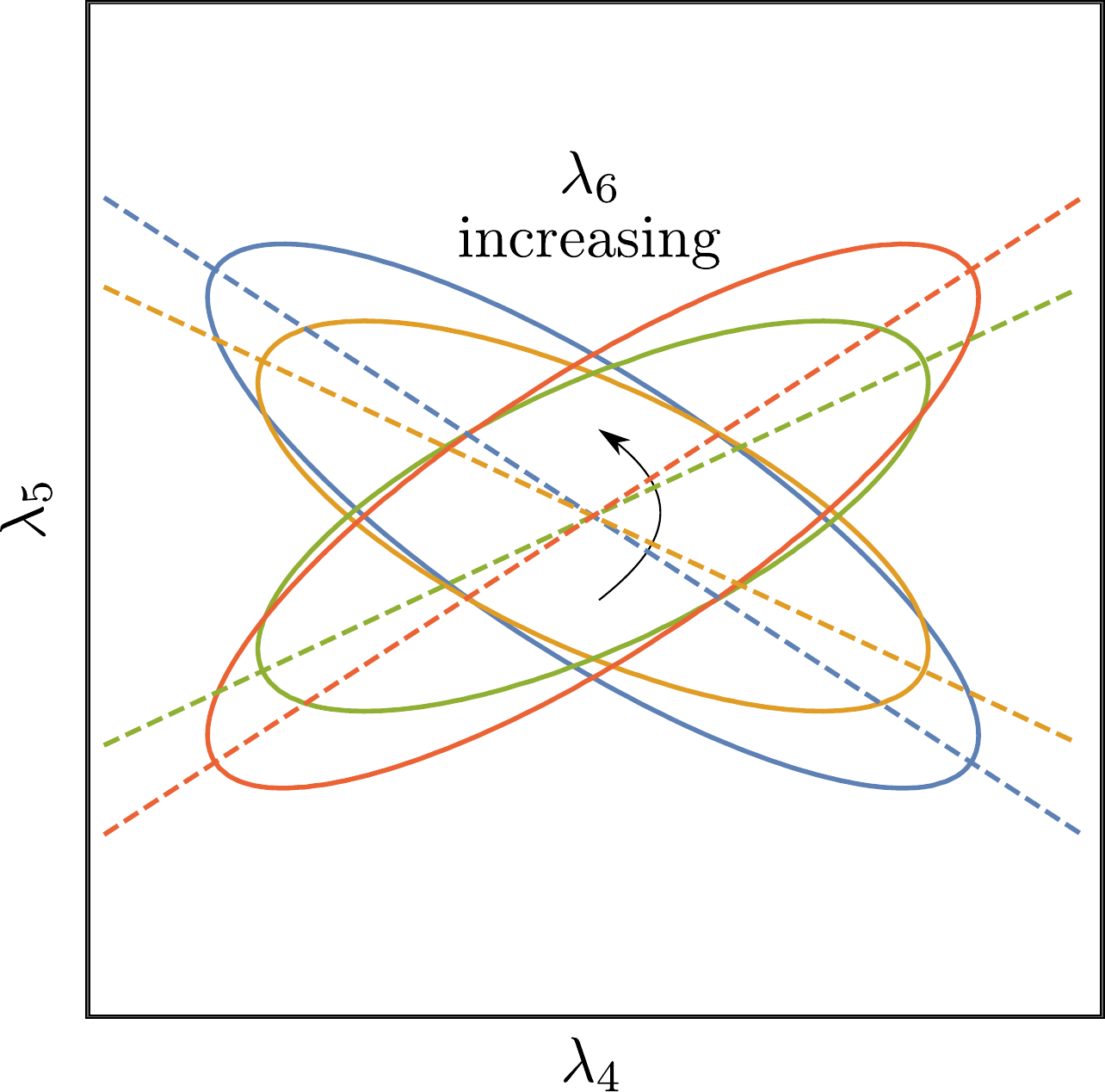}
\end{subfigure}%
\begin{subfigure}{.5\textwidth}
  \centering
  \includegraphics[width=.8\linewidth]{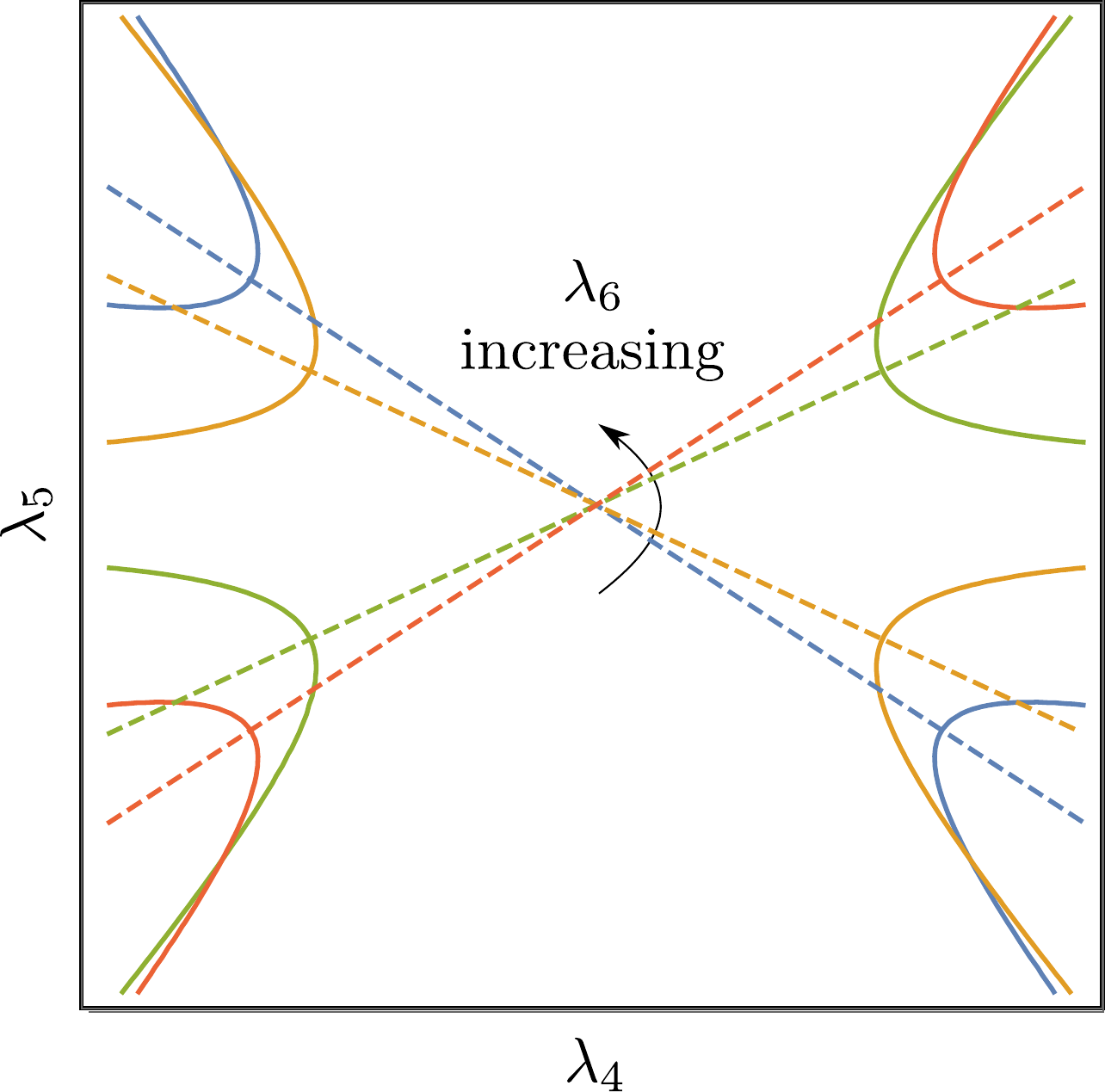}
\end{subfigure}
\caption{Constraints for the parameter space coming from the mass of the Higgs boson produces ellipses and hyperbolas in the $\lambda_5-\lambda_4$ plane according to Eq. (\ref{eq:40}). In this particular example, we took $\lambda_1=4+\lambda_2$, $\lambda_3=3$, the usual values for the VEVs and $m_h=125.18$ GeV. Notice how the conics rotate counter clockwise as $\lambda_6$ increases, c.f. Eq. (\ref{eq:41}).}
\label{conic}
\end{figure}

\begin{figure}[h!]
\centering
\includegraphics[width=.45\linewidth]{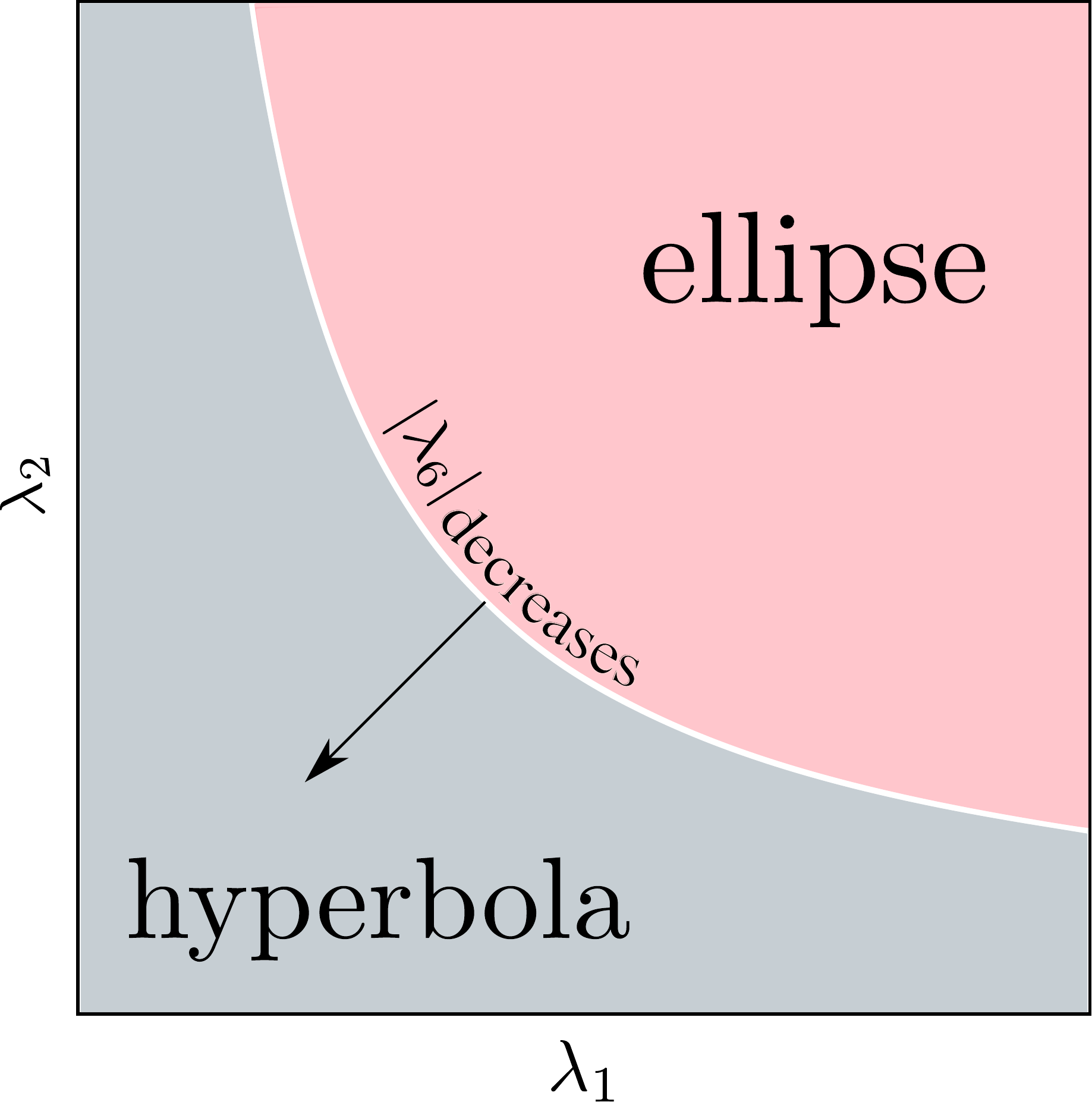}  
\caption{The region separating ellipses and hyperbolas for a $\lambda_3$ fixed and $\lambda_6$ varying.}  
\label{coniccondition}
\end{figure}

Let's return to the square masses of the CP-even scalars, $h$, $H_2$ and $H_3$. Their square masses are obtained from the eigenvalues of the matrix $m_{ij} = \frac{1}{2}\frac{\partial^2 V}{\partial \phi_i \partial \phi_j} \bigg|_{\phi=\textrm{min}},$ where $\phi_i$, $i=1,2,3$, are the real parts of the fields $\eta_1^0$, $\rho_2^0$ and $\chi_3^0$, respectively. In that basis, $m_{ij}$ coincides with $\frac{1}{2}H_0$, where $H_0$ is the Hessian matrix given in Eq. \eqref{eq:21}.  A perturbative analysis in powers of $v/v_\chi$ shows that $m^2_h\propto v^2$ and $m^2_{H_{2,3}}\propto v_\chi^2$. 
For this reason, we have that $m^2_{H_{2,3}}\gg m^2_h$ since $v_\chi^2\gg  v^2$. This observation allows us to calculate an analytical expression for $m^2_h$. In order to do so, it is useful to write the characteristic polynomial of  $\frac{1}{2}H_0$, $P$, as
\begin{eqnarray}
\label{eq:38}
P&=&\det\left[m^2\,{\bf{1}}_{3\times 3}-H_0/2\right]=\cdots+\mathcal{C}\,m^2+\cdots,\nonumber\\
&=&m^6-\Tr\left[H_0/2\right]\,m^4+\det\left[H_0/2\right]\left(1/m^2_h+1/m^2_{H_{2}}+1/m^2_{H_{3}}\right)  \,m^2-\det\left[H_0/2\right], \nonumber \\
\end{eqnarray}
where ${\bf{1}}_{3\times 3}$ is the $3\times 3$ identity matrix and $\mathcal{C}$ is the $m^2$ coefficient of $P$ when calculated from $\det\left[m^2\,{\bf{1}}_{3\times 3}-H_0/2\right]$. Now, since $m^2_{H_{2,3}}\gg m^2_h$,  we can write
\begin{eqnarray}
\label{eq:39}
m^2_h&\simeq&\frac{\det\left[H_0/2\right]}{\mathcal{C}},
\end{eqnarray}
The formula for $m^2_h$ in Eq. \eqref{eq:39} is at $2\%$ accuracy. Because $m_h$ is the Higgs boson mass, it must be $m_h=125.18\pm 0.16$ GeV \cite{Aad:2015zhl, Sirunyan:2017exp,aaboud2018measurement,tanabashi2018aps}. This imposes strong constraints on $\lambda_1,\cdots,\lambda_6$ and $|\lambda_{15}|$. For instance, the $\lambda_4$ and $\lambda_5$ values required to get the Higgs boson mass lie on ellipses or hyperbolas. The particular form of the conic appearing in the $\lambda_5-\lambda_4$ plane depends on the other $\lambda$ values given by Eq. \eqref{eq:39}. To be more specific, let's write the eigenvalue equation for $m_h^2$
\begin{eqnarray}
\label{eq:40}
&&\left(\lambda_2-\frac{m_h^2}{2 v_\rho^2}\right)\lambda_4^2+\left(\lambda_1-\frac{m_h^2}{2 v_\eta ^2}\right)
\lambda_5^2-\lambda_6\,\lambda_4\,\lambda_5 \nonumber\\
&&+\lambda_3\left(\lambda_6^2-4\lambda_1\lambda_2\right)+m_h^2\left(2\lambda_3\left(\lambda_2/v_\eta^2+\lambda_1/v_\rho^2\right)+(4\lambda_1\lambda_2-\lambda_6^2)/2v_\chi^2\right)=0,
\end{eqnarray} 
where we have used $|\lambda_{15}|\ll v_\eta,\,v_\rho$  for simplicity. Eq \eqref{eq:40} represents an ellipse or a hyperbola centered at the origin and rotated by an angle $\theta$ given by
\begin{eqnarray}
\label{eq:41}
\tan2\theta=\frac{\lambda_6}{\lambda_1-\lambda_2+\frac{1}{2} m_h^2\left(\frac{1}{v_\rho^2}-\frac{1}{v_\eta^2}\right)}.
\end{eqnarray}
Other characteristics such as the size of the conic are dominated by $\lambda_{1,2,3,6}$ and the ratios $m_h^2/v_\rho^2$ and $m_h^2/v_\eta^2$. Some particular cases are shown in Fig. \ref{conic} and the region separating ellipses and hyperbolas in the plane $\lambda_2 - \lambda_{1}$ is shown in Fig.  \ref{coniccondition}.

Similar conclusions are obtained for the planes $\lambda_6 - \lambda_{4}$ and $\lambda_6 - \lambda_{5}$. Now, for the most general case, i.e. when $|\lambda_{15}|$ is not negligible, Eq.  \eqref{eq:40} takes the most general form $a\,\lambda_4^2+ 2 b\,\lambda_4\,\lambda_5 + c\,\lambda_5^2+2d\,\lambda_4+2f\,\lambda_5+g = 0$  where the coefficients $a$,$\cdots$,$g$ depend on $\lambda_{1,2,3,6,15}$. Thus, for the general case the ellipse/hyperbola is not centered at the origin and the rotation angle also acquires a dependence on $|\lambda_{15}|$. We are not going to consider that case in detail.

Finally, we can calculate $m^2_{H_{2,3}}$ as a function of $m^2_h$ as follows
\begin{equation}
\label{eq:42}
m^2_{H_{2,3}} = \frac{1}{2}\Tr\left[H_0/2\right]- \frac{1}{2}m^2_h \mp  \frac{1}{2}\sqrt{\left( \Tr\left[H_0/2\right]- m^2_h \right) \left( \Tr\left[H_0/2\right]+ 3 m^2_h \right) + 4 \mathcal{C} }.
\end{equation}
Because we are working in the approximation of $m_{H_{2,3}}>m_h$ and $m_h=125.18\pm 0.16$ GeV \cite{Aad:2015zhl, Sirunyan:2017exp,aaboud2018measurement,tanabashi2018aps}, the positivity of $m^2_{H_{2,3}}$ does not bring any additional constraint on the $\lambda$'s. Also, note that the positivity of the masses of CP-even scalars implies that $\det\, H_0>0$ as required by the last condition in Eq. \eqref{eq:22}.


\section{Conclusions\label{conclusions}}
In this work, we find tree level constraints on the scalar potential couplings of the economical $3-3-1$ model when considerations of vacuum stability and positivity of the squared scalar masses are taken into account. In particular, we consider the model with a discrete $\ZZ_2$ symmetry acting on $\chi$, $u_{4R}$, $\,d{}_{\left(4,5\right)R}$ fields in a non trivial way. Besides all the appealing features discussed in Sec. \ref{model}, this discrete symmetry makes the quartic terms in the scalar potential to have a biquadratic form of the norm of the fields. This allows us to apply copositivity criterion in order to guarantee that the scalar potential is bounded from below. When copositivity criterion is imposed in combination with the first and second derivative tests for the vacuum expectation values given in Eq. \eqref{eq:12}, ten of the scalar couplings are constrained. In more detail, $\lambda_{1,2,3}$ need to be positive and the $\mu_{1,2,3}$ parameters are completely determined, c.f. Eq. \eqref{eq:20}.  Besides that, $\lambda_{4,5,6}$ couplings are constrained from below by the copositivity and from above by the positivity of the principal minors of the Hessian matrix, Eq. \eqref{eq:29}. More interestingly, there is always an excluded region for $\lambda_{4,5,6}$ which we called the minimum excluded region in Figs. \ref{maxminreg4} and \ref{L56}, respectively. This region comes from the copositivity criterion and gives a lower bound for $\lambda_{4,5,6}$. It is remarkable that the excluded region for these $\lambda$ couplings also has a maximum provided all copositivity conditions are satisfied. $\lambda_{7,8,9,10}$ play an important role in determining the form of both the minimum and the maximum excluded regions for  $\lambda_{4,5,6}$. On the other hand, copositivity does not have anything to say about the upper bound on $\lambda_{4,5,6}$ and it is here that second derivative test is important. We analyse the role that $|\lambda_{15}|$ (where we have applied a phase shift in the fields to make $\lambda_{15}$ real and positive without loss of generality) has in determining that upper bound. As the $|\lambda_{15}|$ coupling is technically small, we have studied the bound when $|\lambda_{15}|\ll v_\eta,v_\rho$ showing that the smallest upper bound on $\lambda_{4,5,6}$ is always larger than $2\sqrt{\lambda_1\lambda_3}$, $2\sqrt{\lambda_2\lambda_3}$, $2\sqrt{\lambda_1\lambda_2}$, respectively.

In order to constrain the rest of $\lambda$ couplings, we turn our attention on positivity of the squared scalar masses. After finding general expressions for the masses of the charged and 
CP-odd scalars, we find constraints on $\lambda_{7,8,9}$ given in Eqs. (\ref{eq:32}-\ref{eq:33}), respectively. Actually, we apply a stronger limit for the case of the masses of charged scalars since these, roughly speaking, must be heavier than $80$ GeV. As the constraints on $\lambda_{7,8,9}$ strongly depend on the VEVs, even in the case of $|\lambda_{15}|\rightarrow0$, we estimate the lower bounds on $\lambda_{7,8,9}$ using the VEVs  that satisfy the upper bound on the mixing angle between the two neutral gauge bosons in the model and the lower bound on the mass of $Z'$ gauge boson. Doing that, we obtain the lower bounds in Eq. \eqref{eq:37}. 

Moreover, we find approximate formulas for the squared masses of the CP-even scalars  in the model. If the $v_\rho,v_\eta\ll v_\chi$ hierarchy is satisfied (as assumed in this $3-3-1$ model), the squared masses of the CP-even scalars different from the Higgs boson are proportional to $v_\chi^2$, which allows us to find a $2\%$ accurate formula for the Higgs squared mass, c.f. Eq. \eqref{eq:39}. Using the fact that the Higgs mass must be $m_h=125.18\pm 0.16$ GeV \cite{Aad:2015zhl, Sirunyan:2017exp,aaboud2018measurement,tanabashi2018aps}, we find that $\lambda_5-\lambda_4$, $\lambda_6-\lambda_4$ and $\lambda_6-\lambda_5$ satisfy the ellipse or the hyperbola general equations with coefficients determined by $\lambda_{1,2,3,6,15}$ couplings. We outline the behaviour of such conics for the case $|\lambda_{15}|\ll v_\eta,v_\rho$ in Fig. \ref{conic}. In that case, they are centered at the origin and their rotation angle strongly depends on $\lambda_6$. Furthermore, we find equations for squared masses of the other two CP-even scalars in terms of $m_h^2$ and the trace and determinant of the Hessian matrix, see Eq. (\ref{eq:42}). Because these squared masses are larger than $m_h^2$ in our approach, their positivity do not bring new constraints on $\lambda$ couplings.      

Although the objective of this paper is to derive tree level conditions for the quartic couplings of the scalar potential coming from vacuum stability, the minimization of the scalar potential, the positivity of the squared masses of the extra scalars, the Higgs boson mass, the Z${}^\prime$ gauge boson mass and its mixing angle with the SM Z boson in order to restrict the parameter space, it is also interesting to comment some modifications coming from the running of the coupling constants. Roughly speaking, we expect that major differences when compared with the SM are due to the presence of new particles, such as heavy quarks and leptons. For example, there are three additional quarks (an up-type quark and two down-type quarks) which are assumed to have masses in the scale of the $\textrm{SU}(3)_L$ symmetry breaking, i.e. $1-10$ TeV. For this reason, we expect that these quarks modify significantly the beta functions of the quartic couplings that involve the $\chi$ scalar triplet because the $\ZZ_2$ symmetry acting on $\chi$, $u_{4R}$ and $d_{4,5R}$ fields makes these new quarks gain masses mainly through the $\chi$ scalar triplet. Thus, the beta functions of $\lambda_{4,5}$ are expected to receive large contributions coming from the diagrams where the new quarks are running. However, other $\lambda$ couplings can receive some contributions from these quarks due to the mixings between scalar mass eigenstates. Moreover, we expect that the beta function of the $\lambda$ quartic couplings receive several positive contributions from the one-loop diagrams with scalars running in them. These positive contributions, roughly speaking, will reduce the allowed regions of the $\lambda$ couplings in a similar way as in the triplet and inverse seesaw models \cite{PhysRevD.92.075028,Bonilla:2015kna}. Nevertheless, in order to give a quantitative answer to this question all coupled one-loop renormalization group equations must be carefully studied for this model.


\begin{acknowledgements}
The authors are thankful for the support of FAPESP funding Grant No. 2014/19164-6. B.L.S.V. also thanks DRCC/IFGW at UNICAMP for their kind hospitality. G.G. is supported by CNPq Grant No. 141699/2016-7. C.E.A.S is grateful for the financial support of CNPq, under grant 159237/2015-7, and to the Abdus Salam International Centre for Theoretical Physics for its kind hospitality. The authors thank Renato M. Fonseca and Ana R. Romero Castellanos for useful discussions about the manuscript.
\end{acknowledgements}


\bibliographystyle{unsrt}
\bibliography{References}

\end{document}